\documentclass[a4paper,10pt,DIV=12,flalign, abstracton]{scrartcl}

\setkomafont{disposition}{\fontfamily{ptm}}
\setkomafont{title}{\fontfamily{ptm}}

\usepackage[utf8]{inputenc}
\usepackage[T1]{fontenc}
\usepackage{amssymb,amsmath,amsthm}
\usepackage{nicefrac}
\usepackage{graphicx}
\usepackage{hyperref}
\usepackage{csquotes}
\usepackage{color}
\usepackage{subfig}

 \usepackage[sort&compress,numbers]{natbib}

\newcommand{\stresscomp}{\sigma}
\newcommand{\straincomp}{\varepsilon}
\newcommand{\displcomp}{u}
\newcommand{\bulkmod}{K}
\newcommand{\shearmod}{G}
\newcommand{\emod}{E}
\newcommand{\Poissrat}{\nu}

\newcommand{\bulkmodeff}[1][{}]{\bulkmod_{\mathrm{eff}}^{#1}}
\newcommand{\shearmodeff}[1][{}]{\shearmod_{\mathrm{eff}}^{#1}}
\newcommand{\emodeff}{\emod_{\mathrm{eff}}}
\newcommand{\GibsoncoeffE}{\alpha_{\emod}}
\newcommand{\GibsoncoeffG}{\alpha_{\shearmod}}

\newcommand{\diamcell}{d}
\newcommand{\hexrad}{l}
\newcommand{\wallthickness}{t}
\newcommand{\VVF}{c}
\newcommand{\roundrad}{r}

\newcommand{\height}{H}
\newcommand{\width}{L}
\newcommand{\shearamp}{\gamma}

\newcommand{\shearmodapp}{\shearmod_{\mathrm{app}}}

\newcommand{\strainamp}{\varepsilon}
\newcommand{\emodapp}{\emod_{\mathrm{app}}}

\newcommand{\curvature}{\kappa}
\newcommand{\bendmoment}{M}
\newcommand{\bendstiffness}{K}
\newcommand{\bendstiffnessclass}{K_{\mathrm{class}}}

\newcommand{\stressgradcomp}{R}

\newcommand{\microdisplcomp}{\Phi}
\newcommand{\complmatSGcomp}{B}
\newcommand{\complmatSG}{\hat{\underbar{\complmatSGcomp}}}
\newcommand{\charlengthSGtens}{\ell}

\newcommand{\inputsvg}[1]{\includegraphics{#1}}
\newcommand{\inputgnuplot}[1]{\includegraphics{#1}}

\title{Influence of Topology and Porosity on Size Effects in Stripes of Cellular Material with Honeycomb Structure under Shear, Tension and Bending}

\author{Rinh Dinh Pham, Geralf Hütter\thanks{Technische Universität Bergakademie Freiberg, Institute of Mechanics and Fluid Dynamics, Lampadiusstr.~4, 09596 Freiberg, Germany, Geralf.Huetter@imfd.tu-freiberg.de}
}

\begin{document}

\maketitle

\begin{abstract}

Cellular solids are known to exhibit size effects, i.~e.\, differences in the apparent effective elastic moduli, when the specimen size becomes comparable to the cell size. The present contribution employs direct numerical simulations (DNS) of the mesostructure to investigate the influences of porosity, shape of pores, and thus material distribution along the struts, and orientation of loading on the size effects and effective moduli of regular {honeycomb structures}. Beam models are compared to continuum models for simple shear, uniaxial loading and pure bending {of strips of finite width}. It is found that the {honeycomb} structure exhibits a considerable anisotropy of the size effects and that {honeycomb} structures with circular pores exhibit considerably stronger size effects than those with hexagonal pores (and thus straight struts).
{Positive (stiffening) size effects are observed under simple shear and negative (softening) size effects under bending and uniaxial loading.
The negative size effects are interpreted in terms of the stress-gradient theory.}

\end{abstract}

\section{Introduction}
\label{sec:introduction}

Cellular solids like foams and, more recently, architectured and additively manufactured meta-materials have a high potential for lightweight constructions and smart devices. The assessment of such structures requires knowledge on the effective macroscopic behavior of these materials. In particular, a relation between effective macroscopic behavior and mesostructure is relevant as the latter can be influenced during manufacturing, or even tailored to specific applications.  

It is well-known that the effective elastic moduli and the effective strength depend on mesoscopic topological properties of foams like connectivity of struts (bending/tension dominated) \citep{GibsonAshby,Yoder2019,Soyarslan2019}, distribution of material between struts and nodes \citep{Gong2005,Jang2008,Storm2013,Storm2019} as well as stochastic properties \citep{Jang2008,Tekoglu2011,Redenbach2012,Janicke2013,Liebenstein2018,Muehlich2020}.

Furthermore, considerable size effects have been observed in the behavior of cellular materials, both in elastic and plastic regime, e.~g.\ \cite{Andrews2001,Chen2002,Dillard2006,Liebold2016a,Yang1982,Waseem2013,Rueger2016,Rosi2019,Rizzi2019a}. This means that small specimens behave differently than would be anticipated from sufficiently large specimens using the classical Cauchy theory of continuum mechanics.
Size effects can be described by generalized continuum theories, starting from \citet{Gauthier1975} and \citet{Yang1982} up to a large number of recent works \citep{Dillard2006,Haard2009,Liu2009,Janicke2013,Waseem2013,Wheel2015,Rueger2016,Ha2016,Rizzi2019,Liebold2016a,Liebenstein2018,Huetter2019,Rosi2019,Huetter2020,Kaiser2020}.
Alternatively, direct numerical simulations (DNS) with discretely resolved mesostructures of foams have been performed, firstly in order to investigate the reasons and mechanisms of size effects and, secondly, as benchmark for generalized continuum theories. Most DNS employed beam lattices \citep{Tekoglu2011,Diebels2014,Ha2016,Liebenstein2018,Yoder2019,Yoder2019a,Glaesener2019,Rizzi2019a,Muehlich2020}, but also continuum models have been used \citep{Janicke2013,Wheel2015,Iltchev2015,Ameen2018,Huetter2019,Rosi2019}. The quantitative comparison of DNS of tension tests and shearing tests in \citep{Huetter2019,Huetter2020} exhibited differences, even in normalized data for hexagonal structures between beam models and continuum models.

The scope of the present contribution is to identify the reason for these differences. For this purpose, a systematic study of the effect of distribution of material in struts and alignment of struts in regular hexagonal microstructures on size effects and effective moduli in simple shearing, uniaxial tension and pure bending is performed.

The present paper is structured as follows: Section~\ref{sec:hexagonallattice} summarizes some basic geometric properties of porous materials with hexagonal micro-structure and the employed finite element models, before the subsequent sections~\ref{sec:shear}--\ref{sec:bending} present models and results for simple shear, uniaxial loading and pure bending, respectively. Finally, section~\ref{sec:summary} closes with a summary.

\section{Hexagonal micro-structures}
\label{sec:hexagonallattice}
\begin{figure}%
 \centering
\subfloat[hexagonal pore (prismatic struts)]{\inputsvg{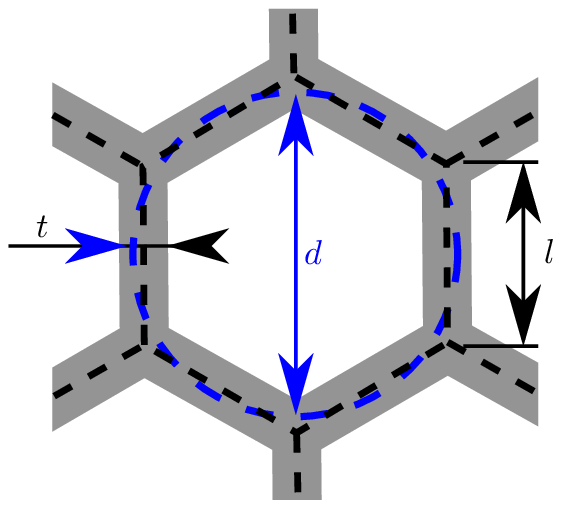}} \hspace{1cm}%
\subfloat[circular pore]{\inputsvg{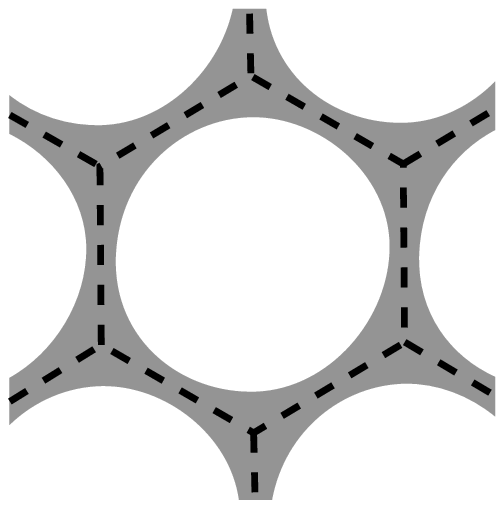}}
\caption{Hexagonal micro-structures}%
\label{fig:hexcell}%
\end{figure}

Cells of a periodic hexagonal micro-structure with respective dimensions are depicted in \figurename~\ref{fig:hexcell}.
For later normalisation of results, the 
diameter $\diamcell=\sqrt{6\sqrt{3}/\pi}\hexrad\approx1.82\,\hexrad$  
of a circle of equivalent area is introduced as characteristic dimension. The porosity $\VVF$ is defined for all configurations as the area fraction $\VVF=A_{\mathrm{pore}}/A_{\mathrm{cell}}$ of pores.

\begin{figure}
	\subfloat[Orientation 1]{\includegraphics[height=0.3\textwidth]{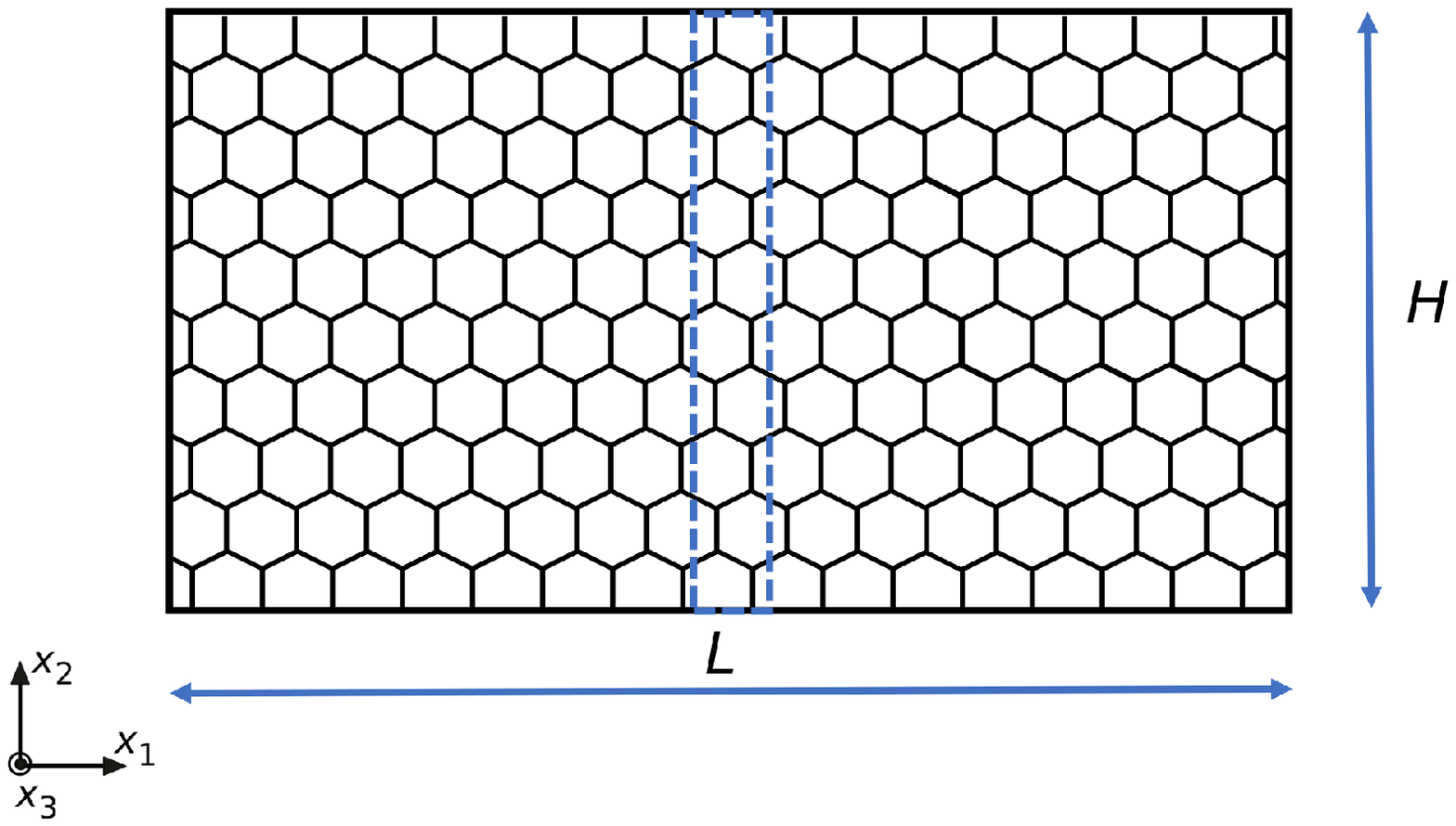}} \hfill
	\subfloat[Orientation 2]{\includegraphics[height=0.3\textwidth]{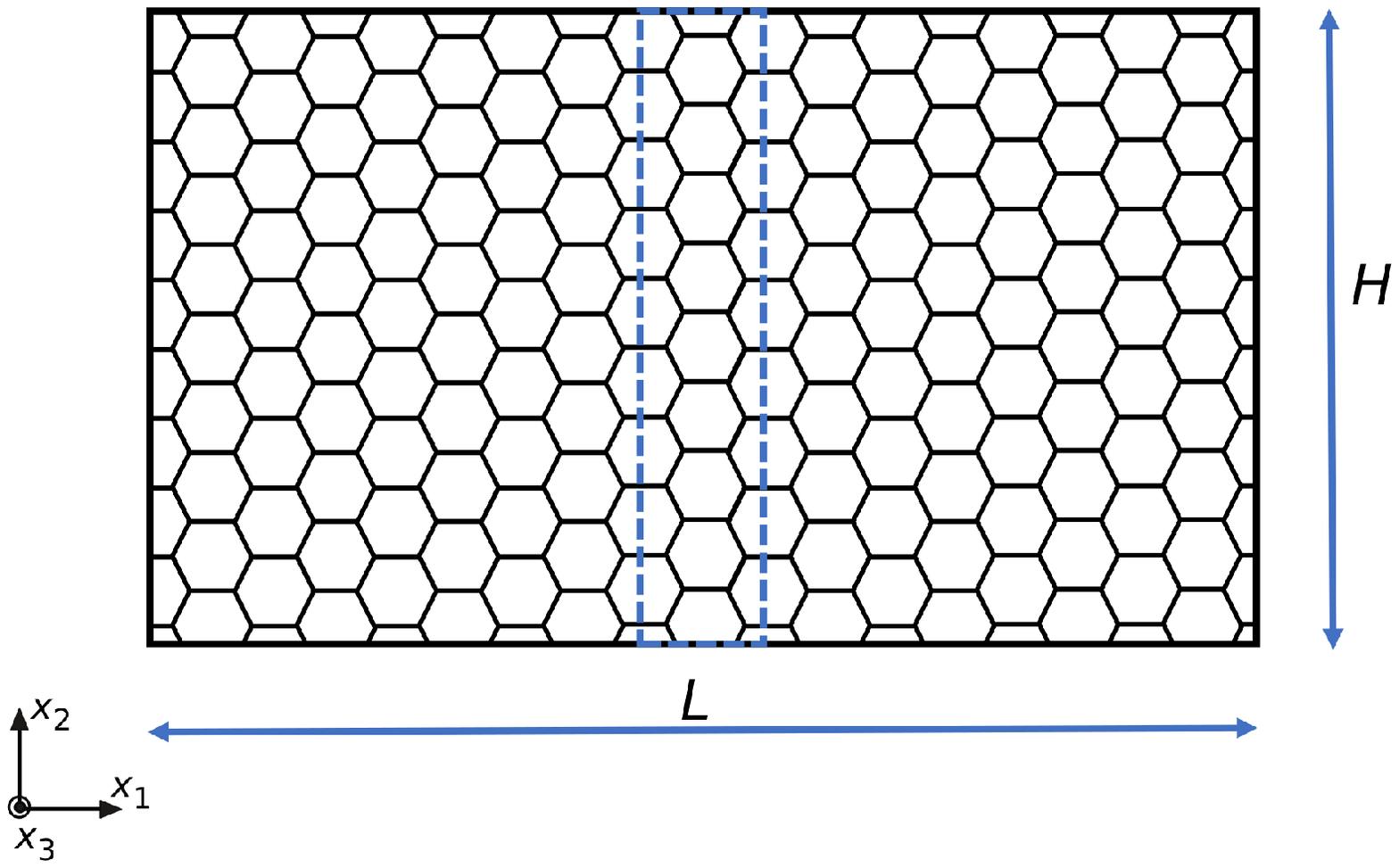}}
 \caption{Cutouts (dashed rectangle) of rectangular specimens for beam models}
 \label{fig:specimencutout}
\end{figure}

In the present study, slender rectangular specimens $\width\times\height$ are simulated under different loading conditions, whose length is assumed to be much larger than height $\width/\height\rightarrow\infty$. Consequently, only a repeatable cutout of the structure is required in the DNS as shown in \figurename~\ref{fig:specimencutout}.  

It is well-known that regular hexagonal structures have isotropic effective in-plain elastic moduli, {see e.g.~\citep{Christensen1987}}. For a hexagonal porous structure with prismatic struts, these effective moduli have been determined by \citet{GibsonAshby} based on beam theory as  
\begin{equation}
	\begin{split}
	\emodeff&=\frac{4}{\sqrt{3}}\emod \left(\frac{\wallthickness}{\hexrad}\right)^3\frac{1}{1+\GibsoncoeffE\,(\wallthickness/\hexrad)^2} \\
	\shearmodeff&=\frac{1}{\sqrt{3}}\emod \left(\frac{\wallthickness}{\hexrad}\right)^3\frac{1}{1+\GibsoncoeffG\,(\wallthickness/\hexrad)^2}\,.
	\end{split}
	\label{eq:effmodGibsonAshby}
\end{equation}
The coefficients $\GibsoncoeffE$ and $\GibsoncoeffG$ depend on whether axial and shear deformations of the struts are considered or neglected and are listed in \tablename~\ref{tab:Gibsoncoeffs}. 
\begin{table}
\centering
\caption{Coefficients for effective moduli in Eqs.~\eqref{eq:effmodGibsonAshby} \citep{GibsonAshby}}
\label{tab:Gibsoncoeffs}
\begin{tabular}{l|c|c}
                             & $\GibsoncoeffE$  & $\GibsoncoeffG$ \\ \hline      
 bending, axial and shear deformation & $5.4+1.5\Poissrat$ & $3.3+1.75\Poissrat$ \\
 bending and axial deformation & $3$ & $1/2$ \\
 only bending        & $0$ & $0$
\end{tabular}
\end{table}
Therein, $\emod$ and $\Poissrat$ refer to Young's modulus and Poisson ratio of the bulk material, respectively. 
However, higher-order effects may still be anistropic. That is why two distinguished orientations are employed in the present study as shown in \figurename~\ref{fig:specimencutout}.

\begin{figure}
	\centering
	\subfloat{\includegraphics[width=0.2\textwidth]{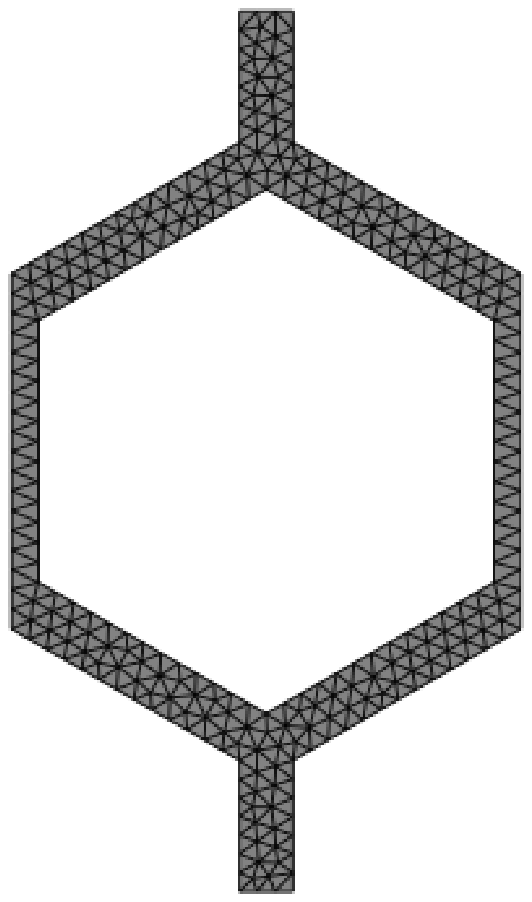}} \hspace{1cm}
	\subfloat{\includegraphics[width=0.2\textwidth]{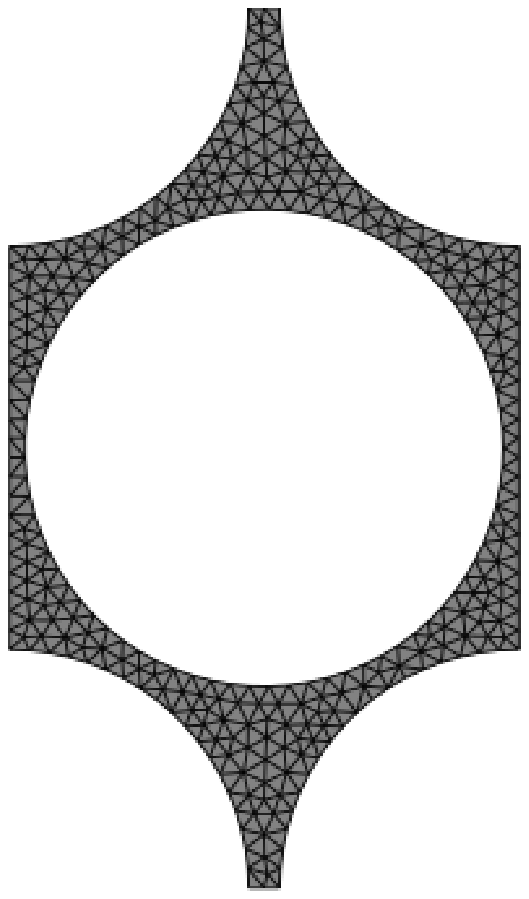}}
	\caption{Coarse finite element meshes of unit cells (the employed meshes are finer)}
	\label{fig:mesh}
\end{figure}

For the finite element simulations, the unit cells have been meshed by triangular linear plane stress elements using the values $\Poissrat=0.3$ as shown in \figurename~\ref{fig:mesh}. After some mesh convergence studies have been performed in advance, it was decided to employ meshes with an element edge length of $\hexrad/800$ to $\hexrad/100$ depending on porosity $\VVF$. Such a mesh resolution is considerably finer than shown in \figurename~\ref{fig:mesh} (unrecognizable if depicted), but is necessary for the structures with high porosity to ensure that there is a number of elements within each cross section of the struts. 
The structures with prismatic struts have been simulated additionally with Timoshenko beam elements.

\section{Simple Shearing}
\label{sec:shear}

\begin{figure}%
\centering
\includegraphics[width=0.45\textwidth]{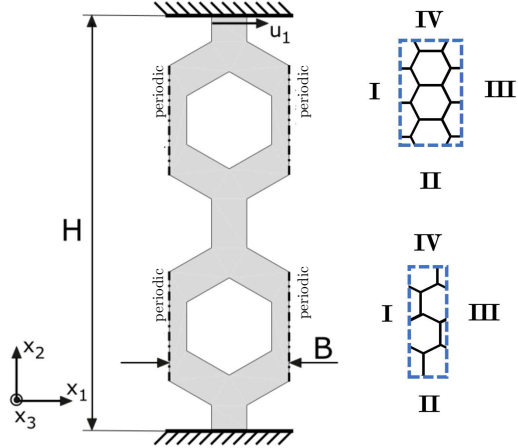}%
\caption{Simple shearing of an infinite layer}%
\label{fig:simpleshear}%
\end{figure}

Firstly, the behavior of an infinite layer under simple shearing is simulated. Due to the periodicity along the layer, only a single slice with thickness of a single unit cell needs to be resolved in the FEM model as depicted in \figurename~\ref{fig:simpleshear}.   
The periodicity conditions at the left and right faces read
\begin{align}
	u_1^I(x_2)&=u_1^{III}(x_2), &
	u_2^I(x_2)&=u_2^{III}(x_2), &
	\varphi_3^I(x_2)&=\varphi_3^{III}(x_2)\,.
\end{align}
Thereby and in the following, the condition with respect to the rotational degree of freedom $\varphi_3$ applies only to beam models. Note that the cut-outs of the beam models are shifted quarter a period in $x_1$-direction compared to those of continuum models since beams must not be located directly at lines of symmetry.  
The boundary conditions at top and bottom read 
\begin{align}
	u_1^{II}(x_1)&=u_2^{II}(x_1)=u_2^{IV}(x_1)=0, &
	u_1^{IV}(x_1)&=\shearamp \height,&
	\varphi_3^{II}(x_1)&=\varphi_3^{IV}(x_1)=0\,,
\end{align}
thus assuming that the rods must not rotate at top and bottom, neither in continuum model nor in beam model.

\begin{figure}
	\subfloat{\includegraphics[scale=0.3]{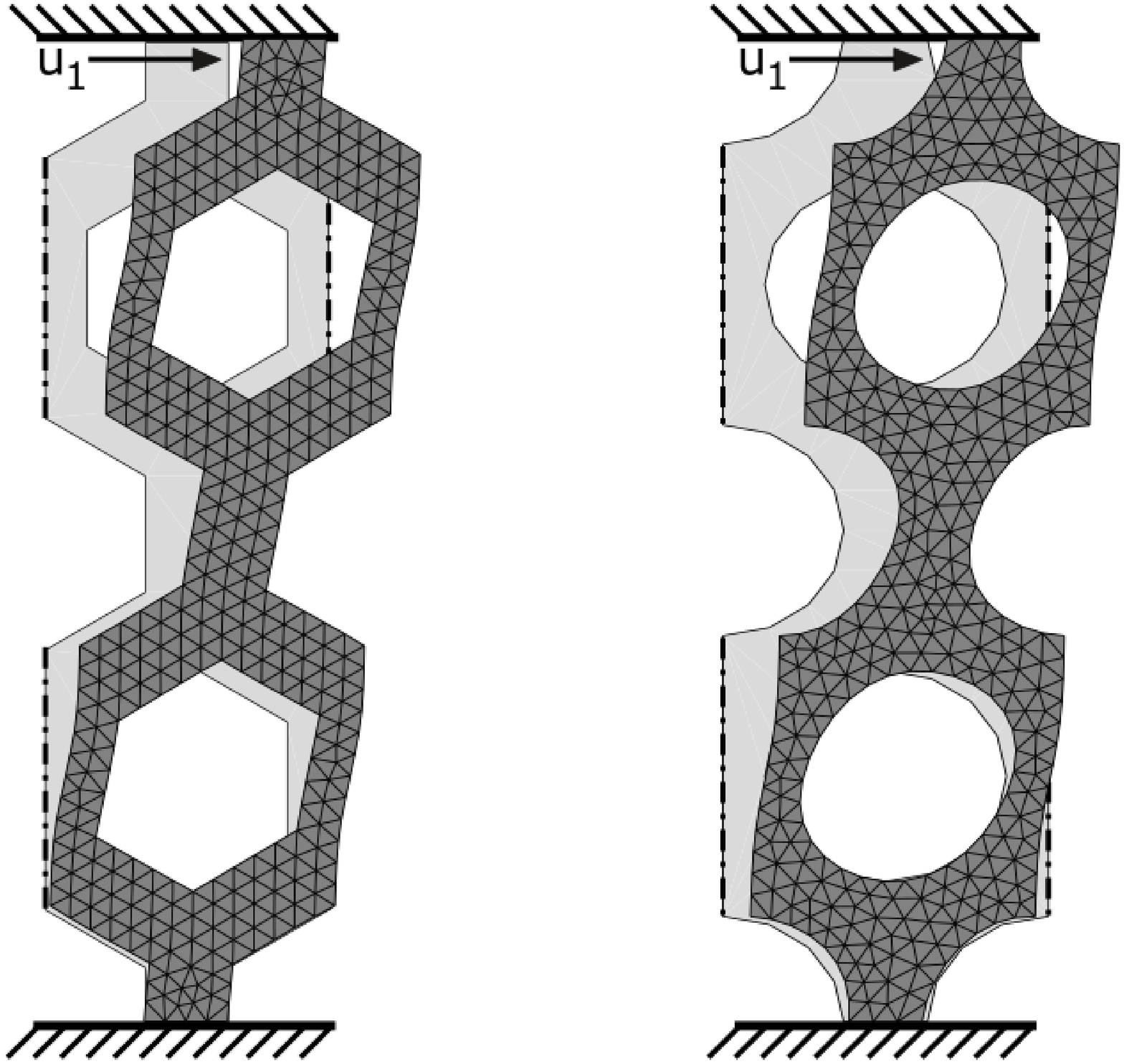}} \hfill 
	\subfloat{\includegraphics[scale=0.3]{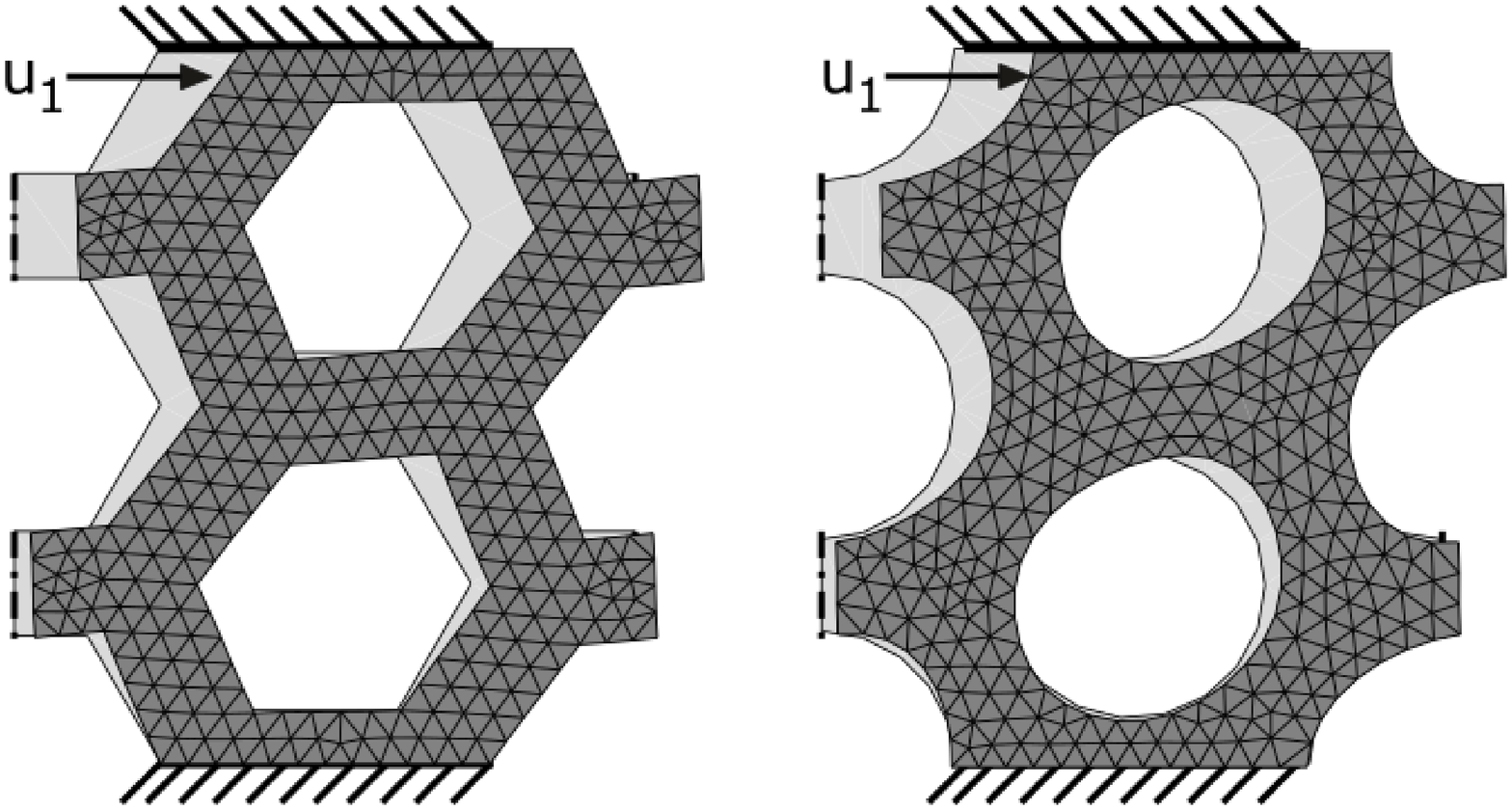}}
	\caption{Local deformation during simple shear with differently aligned microstructures and circular and hexagonal pores for two stacked unit cells ($\VVF=0.5$)}
	\label{fig:defplot_shear}
\end{figure}

Figure~\ref{fig:defplot_shear} shows the deformation of the struts for differently shaped pores and both orientations of the structure relative to the direction of loading.
It can be seen, in particular for orientation 2 at the right-hand side of \figurename~\ref{fig:defplot_shear}, that the rods in the center exhibit an S-shaped deformation, which is not possible at the top and bottom faces. These hindered deformations imply a higher stiffness of the surface layer, which leads to a stiffening size effect, being well-known for simple shear of cellular materials \citep{Tekoglu2008,Tekoglu2011,Janicke2013,Huetter2019}.

In order to quantify the size effect, the apparent shear modulus of each specimen is extracted from the simulations as $\shearmodapp=F_1/(\width\shearamp)$, where $F_1$ denotes the sum of nodal forces at the top surface IV in horizontal direction. Due to linearity of the problem, $\shearmodapp$ is independent of the applied magnitude of loading $\shearamp$.

\begin{figure}
	\centering
	\subfloat[]{
		\includegraphics{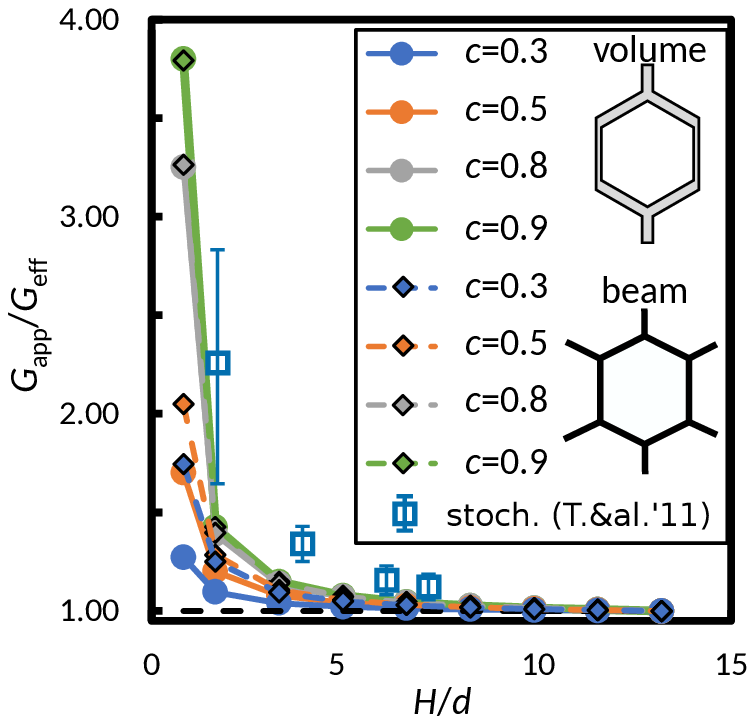}
		\label{fig:sizeeffectshearcmpbeam}}
	\subfloat[]{
		\includegraphics{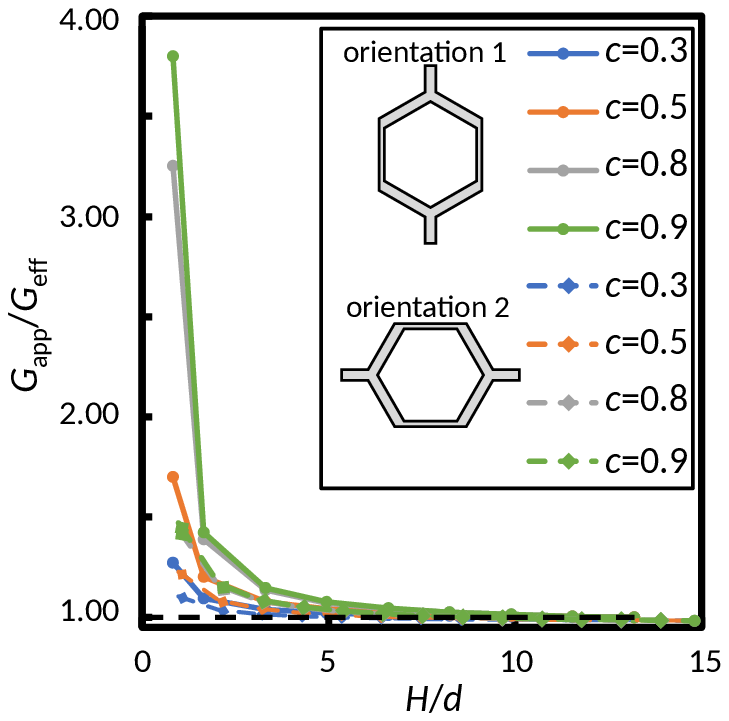}
	\label{fig:sizeeffectshearcmporient}}
	\caption[Apparent shear modulus for simple shear with hexagonal pore]{Apparent shear modulus for simple shear with hexagonal pore: \subref{fig:sizeeffectshearcmpbeam} Comparison of volume model with regular and stochastic beam models (latter results from \citet{Tekoglu2011}) and \subref{fig:sizeeffectshearcmporient} comparison between both orientations}
	\label{fig:sizeeffectshearcmp}
\end{figure}

\begin{figure}
	\centering
	\subfloat[]{\includegraphics{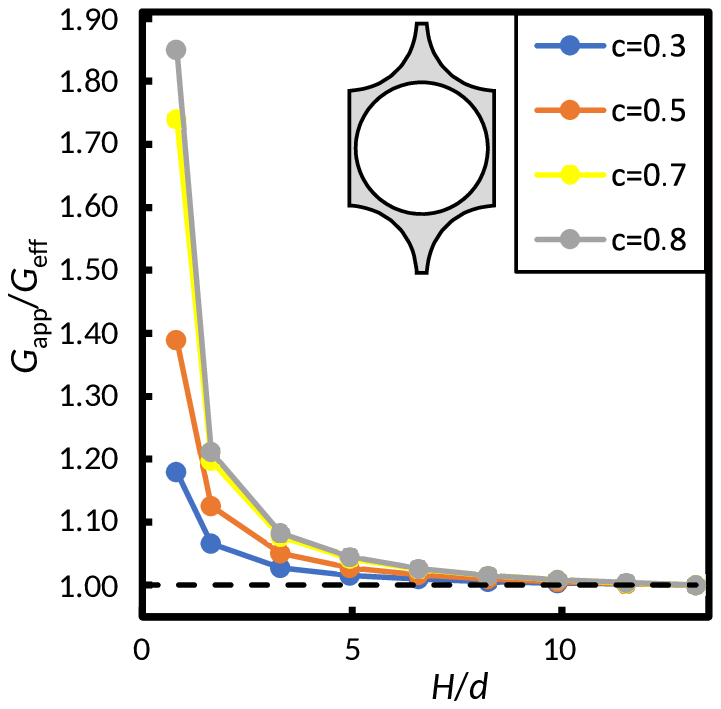}} \hfill
	\subfloat[]{\includegraphics{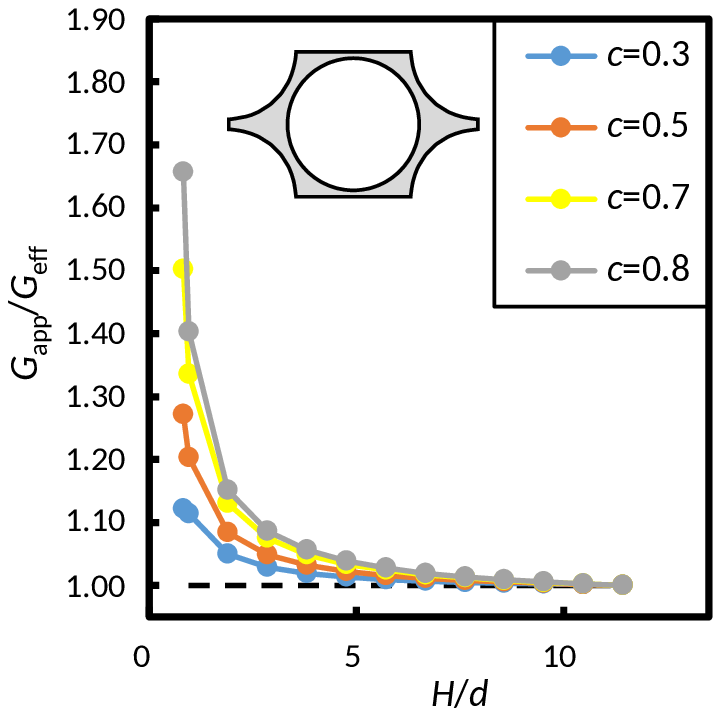}}
	\caption{Apparent shear modulus for simple shear with continuum model with circular pore}
	\label{fig:sizeeffectshearcircpore}
\end{figure}

Figures~\ref{fig:sizeeffectshearcmp} and \ref{fig:sizeeffectshearcircpore} show the extracted apparent shear moduli $\shearmodapp$ for both shapes of pores and for beam models, for both orientations and a number of porosities $\VVF$, each being normalized with the respective effective modulus $\shearmodeff$ of each micro-structure.
The effective shear modulus is defined as 
\begin{equation}
	\shearmodeff=\lim_{\height/\diamcell\rightarrow\infty}\shearmodapp
	\label{eq:effshearmodsimpleshear}
\end{equation}
for each micro-structure, i.~e.\ for each combination of porosity, orientation and shape of pores.
Practically, it is extracted from the simulation with the largest height $\height$ for each material. Consequently, the normalized plots exhibit an asymptote $\shearmodapp/\shearmodeff\rightarrow 1$ (dashed line in the figures).
Firstly, all figures show an increase of the maximum deviation of $\shearmodapp/\shearmodeff$ from 1 with increasing porosity $\VVF$ for all types of models and both orientations. This maximum size effect is obtained for a height of the shear layer of a single unit cell.

Figure~\ref{fig:sizeeffectshearcmpbeam} shows that beam model overestimates the size effect for lower values of porosity $\VVF=0.3$ and $\VVF=0.5$. But for higher porosities $\VVF\ge0.8$, beam model and the volume model with hexagonal pore (and thus prismatic struts) yield virtually identical results as expected. Furthermore, the normalized curves do hardly change anymore when going to even higher porosities $\VVF$, but a saturation of the maximum attainable size effect is reached with respect to $\VVF$. Figure~\ref{fig:sizeeffectshearcmpbeam} incorporates also the results of \citet{Tekoglu2011} with stochastic beam lattices, which exhibit an even stronger size effect than the present regular hexagonal lattices.

Comparing the results for both orientations in \figurename~\ref{fig:sizeeffectshearcmporient}, a strong difference is observed with respect to the maximum attainable size effect, i.~e.\ the maximum value of $\shearmodapp/\shearmodeff$. This means that the size effect is anisotropic, although the classical elastic properties of hexagonal structures are isotropic. The anisotropy of non-classical effects in hexagonal structures has been reported also in \citep{Rosi2019}.

\begin{figure}
	\centering
	\subfloat[]{\includegraphics[width=3cm]{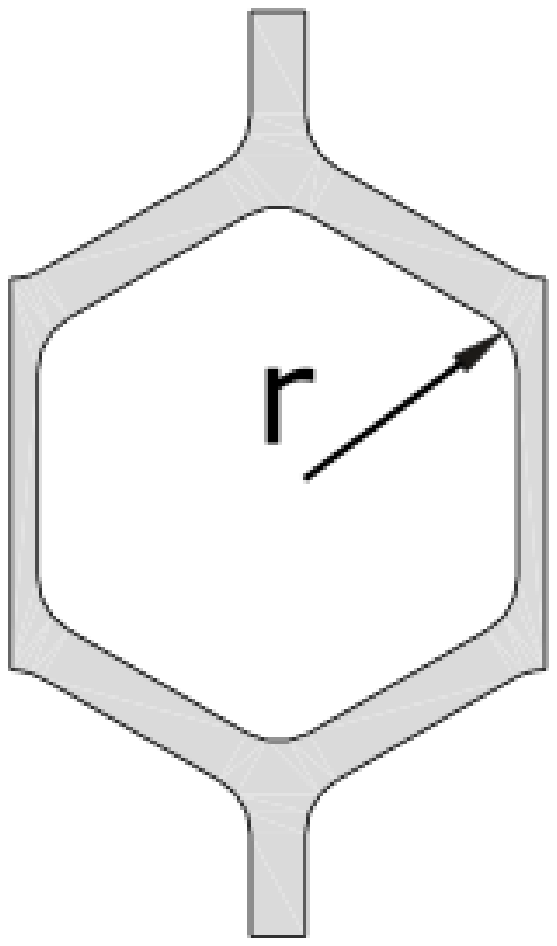}\label{fig:hexcell_rounding}
}\hspace{1cm}
	\subfloat[]{\includegraphics[width=0.45\textwidth]{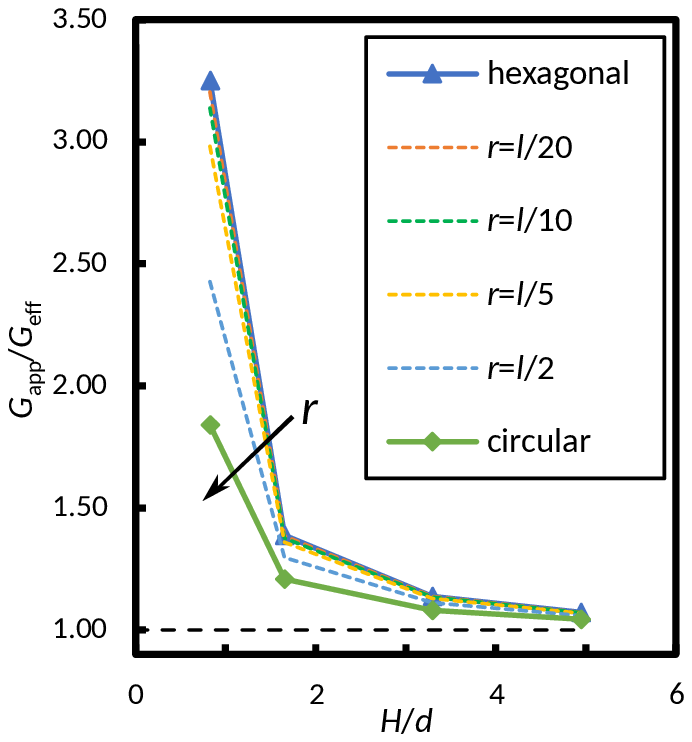}\label{fig:size_effect_shear_rounding_orient1}}
	\caption{Effect of rounding $r$ of pore on size effect in simple shear ($\VVF=0.8$, orientation 1)}
	\label{fig:size_effect_shear_rounding}
\end{figure}

Figure~\ref{fig:sizeeffectshearcircpore} shows the respective data for a hexagonal arrangement of circular pores, which can be simulated by volume models only. The same tendencies are found than with hexagonal pores in \figurename~\ref{fig:sizeeffectshearcmporient}. However, a quantitative comparison with \figurename~\ref{fig:sizeeffectshearcmporient} shows that stronger size effects are obtained with hexagonal pores in orientation 1 than with circular pores and that the difference between both orientations is weaker with circular pores. This behavior is plausible insofar as non-interacting circular pores would be isotropic.
This finding complies with the comparison of results for circular pores with results from literature for hexagonal pores and beam models in \citep{Huetter2019}. To investigate this effect further, models have been created, which posses a rounding $\roundrad$ at the connecting point of the rods as shown in \figurename~\ref{fig:hexcell_rounding}.
The extracted apparent shear moduli in \figurename~\ref{fig:size_effect_shear_rounding_orient1} show a continuous transition from the hexagonal pores ($\roundrad=0$) to the results of circular pores with increasing $\roundrad$.

Note that the maximum attainable size effect $\shearmodapp/\shearmodeff\approx3.9$ with any of the present regular hexagonal foam materials lies considerably below the maximum value $\shearmodapp/\shearmodeff\approx7.5$, which has been reported by \citet{Tekoglu2008} for Voronoi-tesselated stochastic lattices of beams for the same loading conditions. 
Apparently, there is a significant contribution of the stochastic distribution. \Citet{Liebenstein2018} compared stochastic and regular lattice under simple shear. However, they considered shear layers of finite width, where the stiffening size effect from the clamped faces competes  with a softening size effect from unconnected struts at the free lateral faces, making a distinction between both effects impossible.

\begin{figure}
	\centering
	\includegraphics[width=0.7\textwidth]{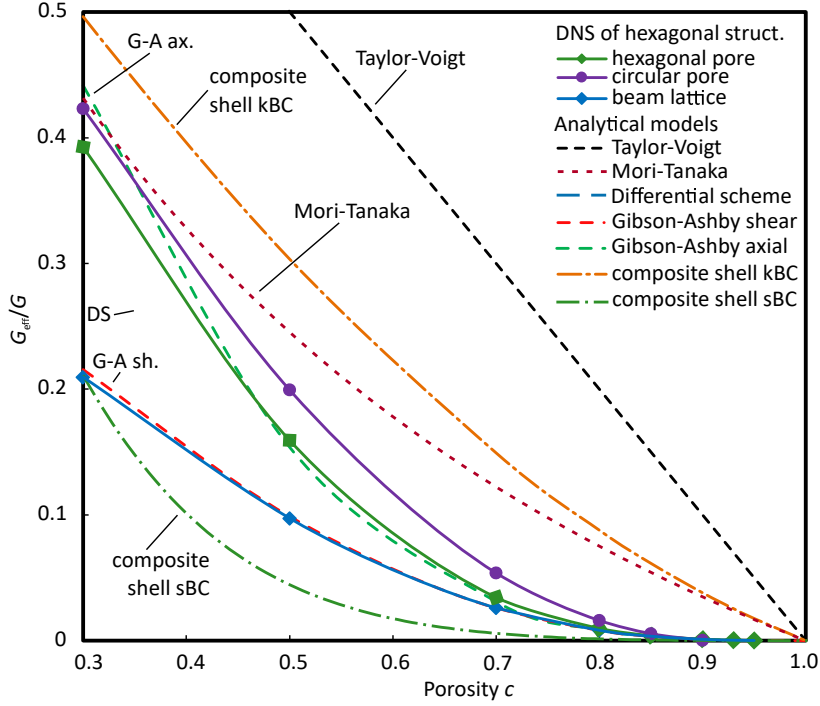}
	\caption{Effective shear moduli from simple shearing tests in comparison with analytical models}
	\label{fig:eff_shearmod}
\end{figure}

Figure~\ref{fig:eff_shearmod} compares the effective shear moduli from the DNS to common analytical models.
Firstly, it is found that the results of the Timoshenko beam models coincide virtually with the analytical solution~\eqref{eq:effmodGibsonAshby} of Gibson and Ashby incorporating shear deformations, thus verifying the present implementation. 
{The present results for the continuum model with hexagonal pores, and thus prismatic struts, comply qualitatively and quantitatively with those of \citet{Lee1996} obtained by a computational unit cell analysis with periodic boundary conditions (not shown here).  }
For high porosities $\VVF\gtrsim0.75$, an agreement of beam model and continuum model with hexagonal pores is observed as expected. For lower porosities, $\VVF\lesssim0.7$, the beam model underestimates the effective shear moduli of the more exact continuum models with hexagonal and circular pores. 
{At same porosity, circular pores lead to higher effective modulus compared to hexagonal pores as predicted qualitatively by \citet{Warren1987}. Though, the model of \citeauthor{Warren1987} overestimates the effective modulus in this regime, cf.~\citet{Lee1996}.}
The relative difference in the effective moduli between both pore geometries decreases with further decreasing $\VVF$ as the interaction between neighboring pores decreases. 
In this regime, the DNS results lie between the estimates of the Mori-Tanaka scheme and differential scheme (both from \citep{Gross2006}). The latter provides a good approximation from high to medium porosities. Remarkably, it is outperformed by the Gibson-Ashby model \eqref{eq:effmodGibsonAshby} incorporating only bending and axial deformations (i.e.\ based on Euler-Bernoulli beam theory). Apparently, the additional compliance due to the shear deformations of the struts is overcompensated by the additional stiffness due to material concentration at the nodes of the structure. The Gibson-Ashby model with bending but no axial deformations (third entry in \tablename~\ref{tab:Gibsoncoeffs}) overestimates the effective modulus considerably (not shown).
The composite shell approach of \citet{Hashin1962}, adopted to the plane case in form of a circular pore in a circular unit cell as outlined in the appendix~\ref{sec:composite shell}, with kinematic or static boundary conditions (\enquote{kBC}/\enquote{sBC}), corresponding to the Hashin-Shtrikman bounds for the present problem, enclose the aforementioned data, but are too far away to be of practical relevance. 
The Taylor-Voigt estimate yields yields an even higher upper bound as expected.

\section{Uniaxial Loading}
\label{sec:uniaxial}

\begin{figure}%
\centering
\includegraphics[width=0.45\textwidth]{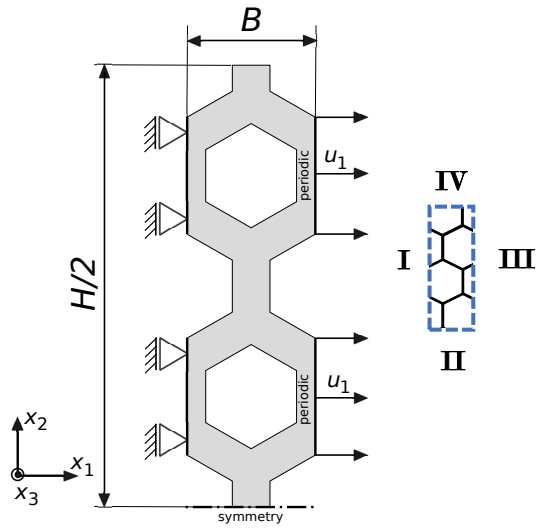}%
\caption{Uniaxial tension of an infinite layer}%
\label{fig:uniaxial}%
\end{figure}

As second load case, the uniaxial tension or compression of an infinitely long specimen is investigated by the simulation of a periodically continuable section as shown in \figurename~\ref{fig:uniaxial}, making additional use of the mirror symmetry.
The essential boundary and periodicity conditions at the four faces read
\begin{align}
	u_1^I(x_2)&=0,& 
	u_1^{III}(x_2)&=\width \strainamp, &
	u_2^I(x_2)&=u_2^{III}(x_2), &
	\varphi_3^I(x_2)&=\varphi_3^{III}(x_2),&
  u_2^{II}(x_1)&=0, &
	\varphi_3^{II}(x_1)&=0\,.
\end{align}
All other degrees of freedom are left unconstrained. Therein, $\strainamp$ denotes the macroscopic longitudinal strain. The apparent Young's modulus is extracted from each simulation as $\emodapp=F_1/(\height\strainamp)$, whereby $F_{1}$ denotes the sum of all reaction forces at the right face III. The effective Young's modulus is defined in analogy to the previous shearing case for each material as $\emodeff=\lim_{\height/\diamcell\rightarrow\infty}\emodapp$. 

\begin{figure}
	\centering
	\includegraphics{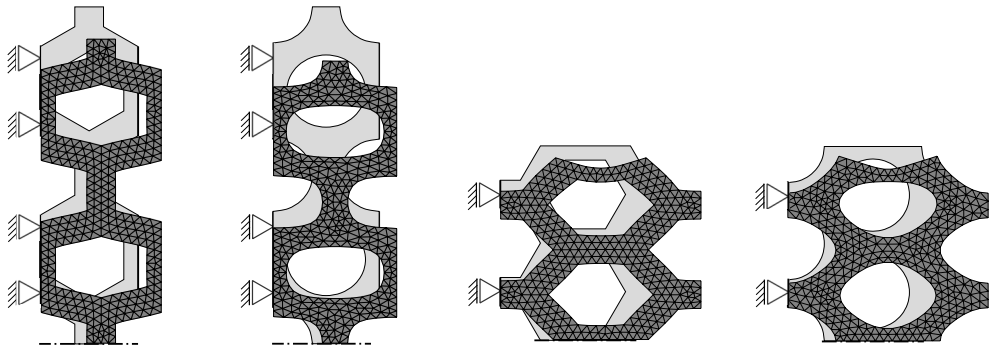}
	\caption{Deformed configurations in uniaxial tension}
	\label{fig:defplot_uniaxial}
\end{figure}

Respective deformed configurations for small specimens with two unit cells in height are shown in \figurename~\ref{fig:defplot_uniaxial}.
It can be seen for orientation 2 at the right-hand side that the struts at the free surface at top get warped, which is not possible for the struts at the line of symmetry at bottom. In contrast, no significant difference in the deformation modes of struts at top and bottom can be observed for orientation 1 at the left-hand side of \figurename~\ref{fig:defplot_uniaxial}.

\begin{figure}
	\centering
	\includegraphics{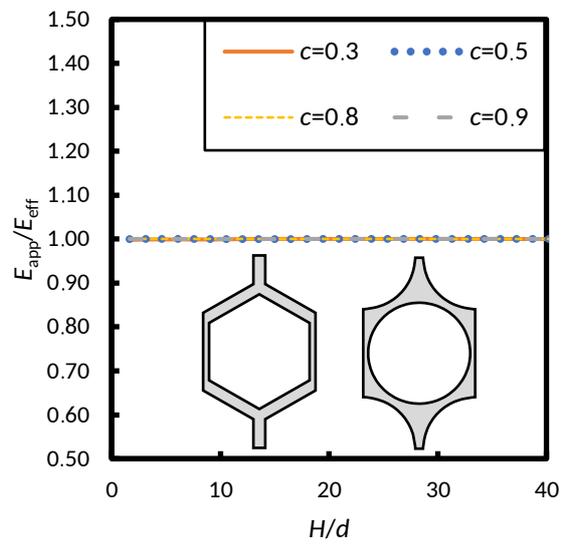}
	\caption{Apparent Young's modulus for uniaxial loading from continuum models with orientation 1}
	\label{fig:sizeeffectuniaxorient1}
\end{figure}

\begin{figure}
	\centering
	\subfloat[]{\includegraphics[height=0.45\textwidth]{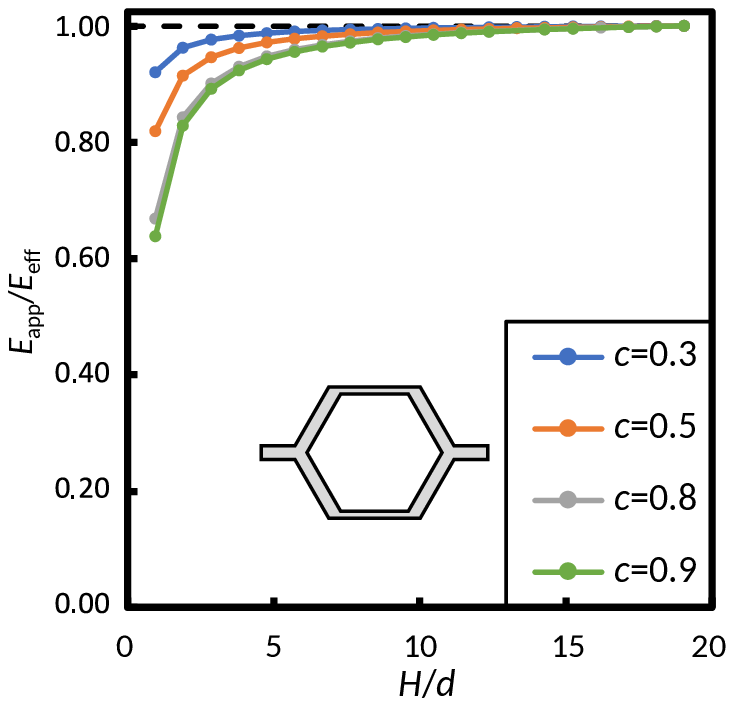}\label{fig:sizeeffectuniaxhexporeorient2}}
	\subfloat[]{\includegraphics[height=0.45\textwidth]{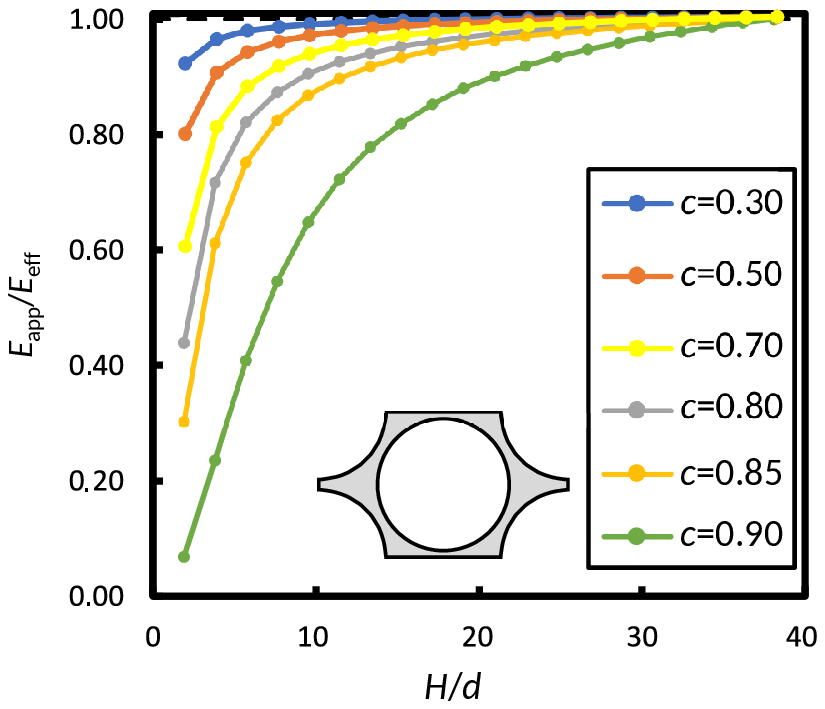}\label{fig:sizeeffectuniaxcircporeorient2}}
	\caption{Apparent Young's modulus for uniaxial loading from continuum models with orientation 2}
	\label{fig:sizeeffectuniaxorient2}
\end{figure}

Consequently, virtually no size effect can be observed for orientation 1 as shown in \figurename~\ref{fig:sizeeffectuniaxorient1}. 
In contrast, \figurename~\ref{fig:sizeeffectuniaxorient2} indicates a considerable softening size effect for orientation 2, i.~e.\ smaller specimens have a smaller apparent Young's modulus than larger ones in accordance with \citep{Tekoglu2011,Liebenstein2018}.
Again, the magnitude of size effect increases with increasing porosity $\VVF$. 
Note that the normalized curves for hexagonal pores in \figurename~\ref{fig:sizeeffectuniaxhexporeorient2} converge for $\VVF\gtrsim0.8$, whereas no such convergence can be observed for the circular pores in \figurename~\ref{fig:sizeeffectuniaxcircporeorient2}. 
Rather, the size effect increases with circular pores up to the maximum porosity $\VVF_{\max}\approx0.907$, whereby the sensitivity with respect to changes of $\VVF$ even increases in this regime. 
Remarkably, such a type of behavior had been predicted for circular pores in \citep{Huetter2020} by a homogenization approach within the stress-gradient theory\footnote{The term \enquote{stress-gradient theory} is not used uniquely in literature, but different non-equivalent formulations have been proposed under this name. Within the present paper, this name refers solely to the formulation of \citet{Forest2012}. } (other shapes of pores were not considered).

Comparing the magnitudes of size effects for simple shear and uniaxial loading in figures~\ref{fig:sizeeffectshearcmp}, \ref{fig:sizeeffectshearcircpore} and \ref{fig:sizeeffectuniaxorient2} shows that a notable size effect in shearing can be observed for specimens $\height/\diamcell\lesssim10$, whereas a notable deviation between $\emodapp$ and $\emodeff$ can be observed with specimens of at least twice that cross section $\height/\diamcell\lesssim10\dots15$. And a saturation of the size effect with increasing porosity was observed in \figurename~\ref{fig:sizeeffectshearcircpore} even with circular pores.

\begin{figure}
	\centering
	\includegraphics{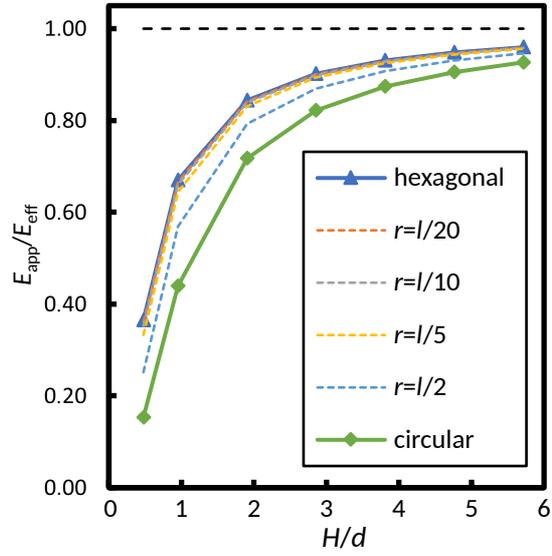}
	\caption{Effect of rounding $r$ of pore on size effect under uniaxial loading ($\VVF=0.8$, orientation 2)}
	\label{fig:size_effect_uniaxial_rounding}
\end{figure}

Figure~\ref{fig:size_effect_uniaxial_rounding} shows the effect of the rounding radius $\roundrad$ of the pore (cmp.\ \figurename~\ref{fig:hexcell_rounding}) on the size effect under uniaxial loading.
It is found that the magnitude of size effect increases with rounding radius. This the opposite trend than observed for simple shear in \figurename~\ref{fig:size_effect_shear_rounding}. However, the effect of $\roundrad$ is weaker under uniaxial loading than for shearing.

\begin{figure*}%
\centering
\includegraphics[width=0.6\columnwidth]{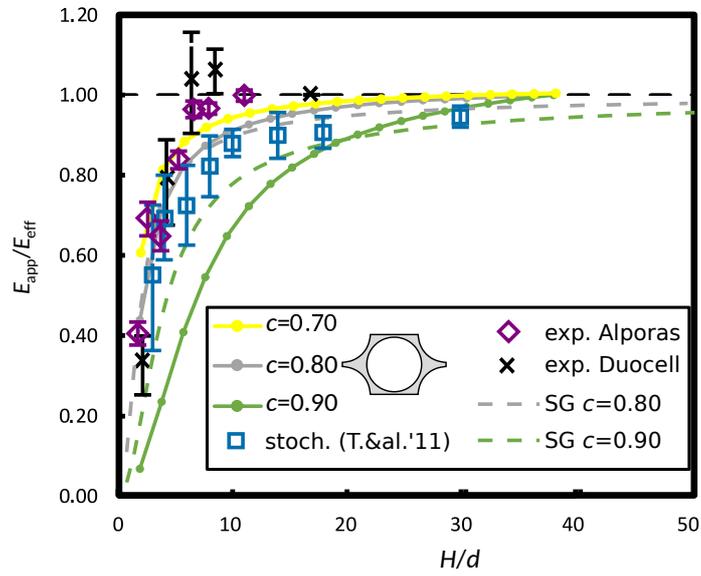}%
\caption{Size effect in apparent Young's modulus of regular hexagonal structure (orientation 2 with circular pore) in comparison to stochastic beam models of \citet{Tekoglu2011}, experimental data of \citet{Andrews2001} and predictions of stress-gradient theory (\enquote{SG}) from \citep{Huetter2020}}%
\label{fig:size_effect_uniaxial_cmp}%
\end{figure*}

Figure~\ref{fig:size_effect_uniaxial_cmp} compares the predictions of the present DNS of regular hexagonal structures with stochastic DNS of \citet{Tekoglu2011}, experimental data of \citet{Andrews2001} and the aforementioned predictions of \citet{Huetter2020} obtained by means of a homogenization approach within the stress-gradient theory.
Both, the present DNS with circular pores and the stochastic beam models of \citet{Tekoglu2011} agree with the experimental data in principle. The data points from the experiments look as if the investigated specimens have not been large enough to reach the asymptotic effective modulus $\emodeff$. Comparing the data of \citet{Tekoglu2011} with the present DNS shows that the stochastic structure exhibits a similar size effect than the present data for regular structures with prismatic struts, in contrast to the previous shearing case. 
The stress-gradient theory gives a very good prediction for $\VVF=0.8$, but underestimates the strong increase of the size effect when going to $\VVF=0.9$. This discrepancy is surely related to the simplified geometry of a circular pore in a circular volume element, which has been obtained for homogenization within the stress-gradient theory.

A plot of the ratio of effective Young's modulus $\emodeff$ and respective matrix value $\emod$ versus the porosity $\VVF$ is qualitatively even almost quantitatively identical to the respective plot for the shear moduli in \figurename~\ref{fig:eff_shearmod} (as the effective Poisson ratio depends only very weakly on $\VVF$) and thus omitted here.

\section{Pure Bending}
\label{sec:bending}

\begin{figure}%
\centering
\includegraphics[width=0.45\textwidth]{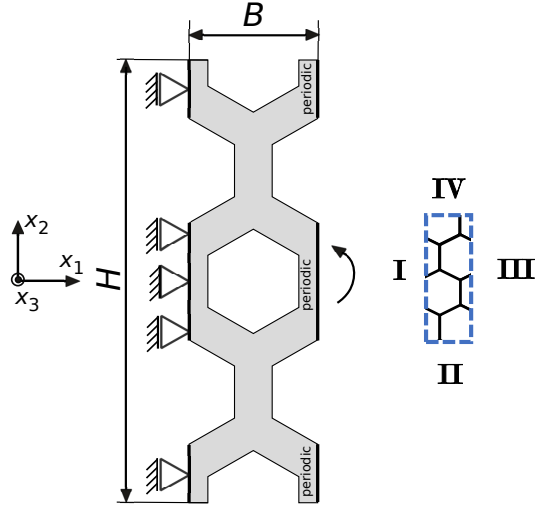}%
\caption{Pure bending of an infinitely long beam}%
\label{fig:bending}%
\end{figure}

The last considered loading case is pure bending of an infinitely long and thus ideally slender beam, implemented as sketched in \figurename~\ref{fig:bending}.
The essential boundary and periodicity conditions at the left and right faces are formulated according to the classical beam theory as
\begin{align}
	u_1^I(x_2)&=0,&
	u_2^I(x_2=0)&=0, &
	u_1^{III}(x_2)&=-\curvature\width x_2,&
	u_2^{III}(x_2)&=u_2^{I}(x_2)+\frac12\curvature\width^2, &
	\varphi_3^{III}(x_2)&=\varphi_3^{I}(x_2)+\curvature\width,&
	\label{eq:BC_bending}
\end{align}
whereby $\curvature$ denotes the mean curvature of the neutral axis $x_2=0$. The top and bottom surfaces are free and are thus left unconstrained (trivial natural boundary conditions). The bending moment is extracted from the simulations  as $\bendmoment=\sum\limits_{i}x_2 F_{1}^{(i)}$ from the horizontal reaction forces $F_{1}^{(i)}$ of all nodes $i$ at the right face. Reaction moments as potential contributions to $\bendmoment$ for beam models turned out to be negligible. The bending stiffness is then defined as $\bendstiffness=\bendmoment/\curvature$ and normalized to the value $\bendstiffnessclass=\emodeff\height^3/12$, which would be expected from the classical theory of bending. Thereby, values of the effective Young's modulus $\emodeff$ from uniaxial loading in Section~\ref{sec:uniaxial} are used. 

\begin{figure}
	\centering
	\includegraphics{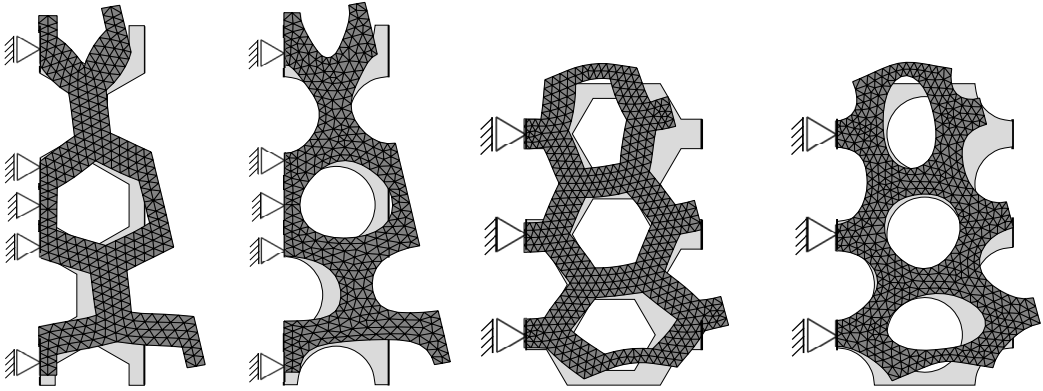}
	\caption{Deformed configurations in pure bending ($\VVF=0.5$)}
	\label{fig:defplot_bending}
\end{figure}

The local deformations under bending are shown in \figurename~\ref{fig:defplot_bending} for both orientations and both shapes of pores.
It can be seen that the unit cells at the free surfaces at bottom and top are deformed similarly than under tension, compare \figurename~\ref{fig:defplot_uniaxial}.

\begin{figure}
	\centering
	\subfloat[Orientation 1]{\includegraphics[width=0.45\textwidth]{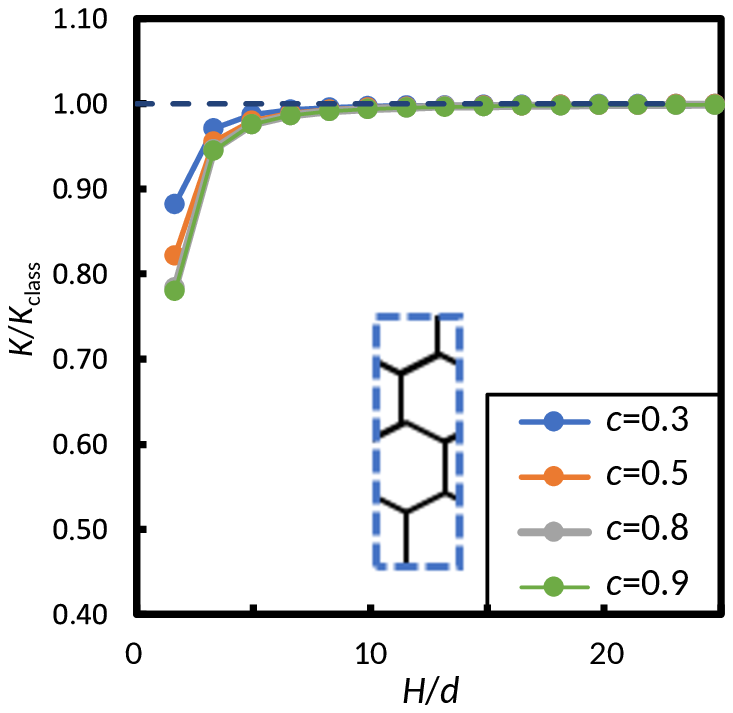}\label{fig:sizeeffectbendingbeam_orient1}}
	\subfloat[Orientation 2]{\includegraphics[width=0.45\textwidth]{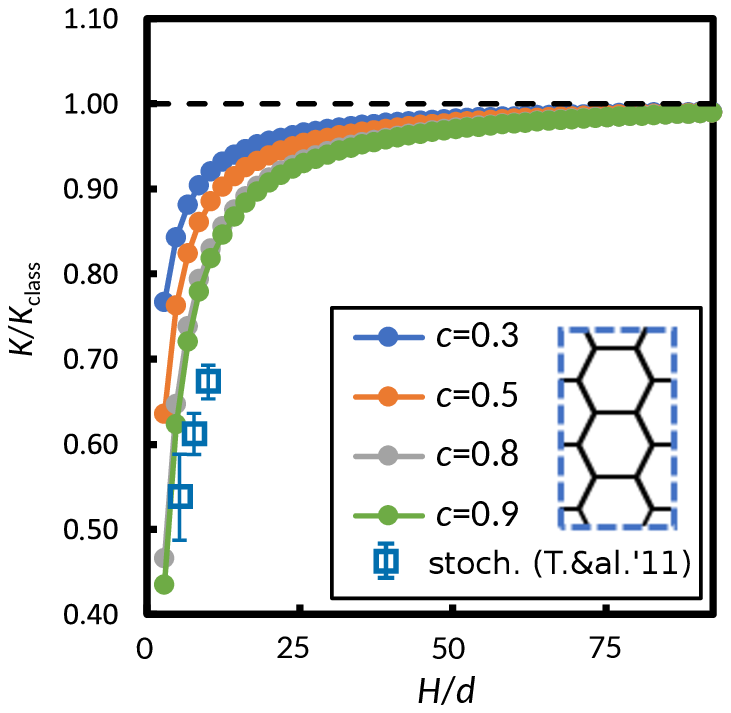}\label{fig:sizeeffectbendingbeam_orient2}}
	\caption{Bending stiffness with beam models for both models in comparison with results for stochastic beam lattices from \citet{Tekoglu2011}}
	\label{fig:sizeeffectbendingbeam}
\end{figure}

Figure~\ref{fig:sizeeffectbendingbeam} shows the normalized bending stiffnesses from the beam models. 
Firstly, it can be seen that all curves approach $\bendstiffness=\bendstiffnessclass$ asymptotically for $\height/\diamcell\rightarrow\infty$ as expected, thus verifying the employed periodicity and boundary conditions~\eqref{eq:BC_bending}. A negative size  effect (smaller specimens are less stiff than expected) is observed for \emph{both} orientations, in contrast to uniaxial loading. Furthermore, the normalized curves converge towards an asymptotic one with increasing porosity $\VVF$ for both orientations. 
Though, the maximum deviation from the classical solution is considerably larger for orientation 2 than for orientation 1.
Having an additional look at the abscissas in \figurename~\ref{fig:sizeeffectbendingbeam_orient1} and \figurename~\ref{fig:sizeeffectbendingbeam_orient2} shows, that a notable size effect is obtained till $\height/\diamcell\lesssim10$ for orientation 1, but even till $\height/\diamcell\lesssim75$ for orientation 2. 
Figure~\ref{fig:sizeeffectbendingbeam_orient2} incorporates additionally the results of \citet{Tekoglu2011} with stochastic beam lattices of high porosity $\VVF>0.9$. It is found that the size effect with stochastic lattice is even slightly larger than the large-porosity curve with regular lattices in orientation 2 as it was found already for the shearing case.

A negative size effect was observed already for orientation 1 in \citep{Wheel2015} and has been attributed to the unconnected struts at the free surface, which do not carry any load and thus form effectively a stress-free surface layer of half a unit cell thickness. This is why the size effect for orientation 1 in \figurename~\ref{fig:sizeeffectbendingbeam_orient1} depends only weakly on the porosity $\VVF$.
Figure~\ref{fig:defplot_bending} shows that the negative size effect for orientation 2 arises rather from a stronger bending deformation of the struts close to the specimen surface. This different mechanism explains the strong effect of $\VVF$ on the size effect in \figurename~\ref{fig:sizeeffectbendingbeam_orient2}.
\Citet{Wheel2015} observed even positive size effects for configurations where the specimen surface consists of continuous matrix material. However, such a configuration can be found in the considered regular hexagonal micro-structures only for low porosities, but not for the present foam-like materials. Though, positive size effects under bending have been also observed experimentally by Lakes et al.~\citep{Yang1982,Rueger2016,Anderson1994a}. They cannot be explained by the present highly idealized model.

\begin{figure*}%
\centering
\includegraphics[width=0.7\columnwidth]{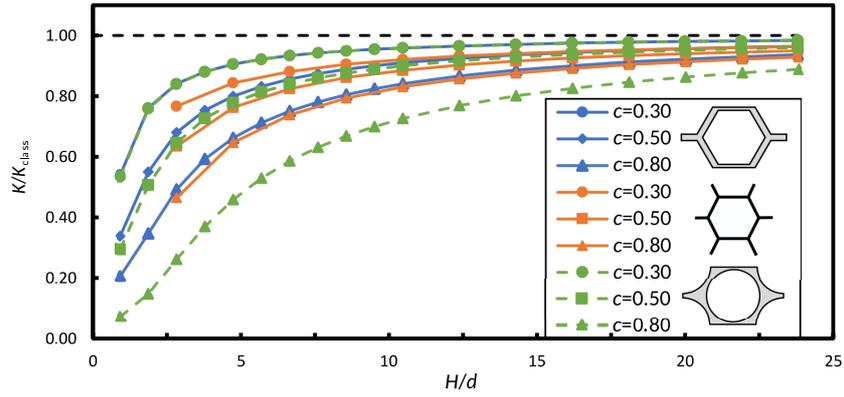}%
\caption{Size effects in bending: Comparison of models (orientation 2)}%
\label{fig:sizeeffectbendingcmp_orient2}%
\end{figure*}

Figure~\ref{fig:sizeeffectbendingcmp_orient2} shows the extracted bending stiffnesses of the different models in normalized form.
It can be seen that the beam model complies to the results of the continuum model with hexagonal pores for sufficiently high porosities $\VVF\gtrsim0.5$. For a medium porosity $\VVF=0.5$, the models with hexagonal and circular pores predict the same size effect. For the higher porosity $\VVF=0.8$, the predicted size effect with circular pores is even stronger than the size effect with hexagonal pores. This is the opposite trend than observed for shearing.

\begin{figure}%
\centering
\subfloat[]{\includegraphics[width=0.8\textwidth]{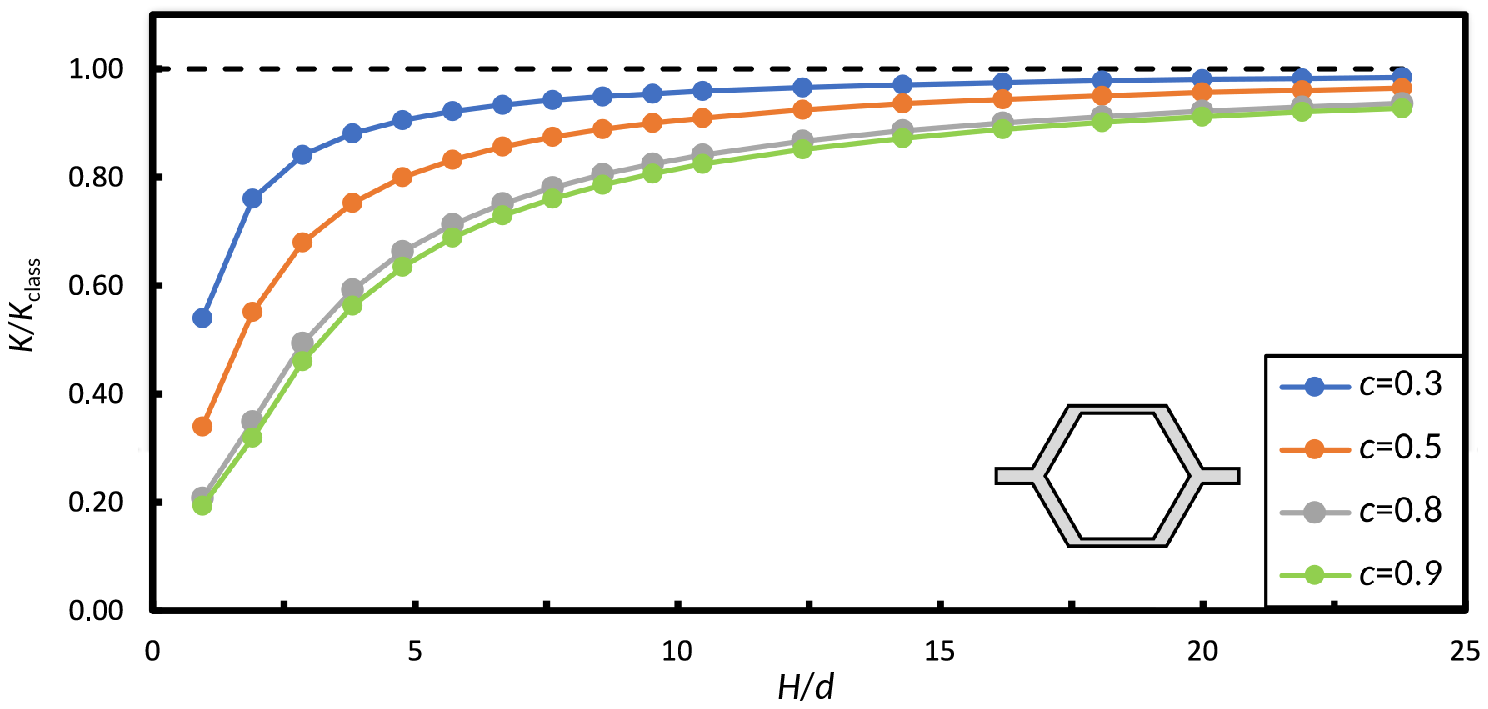}\label{fig:sizeeffectbending_hexpore_orient2}} \\%
\subfloat[]{\includegraphics[width=0.8\textwidth]{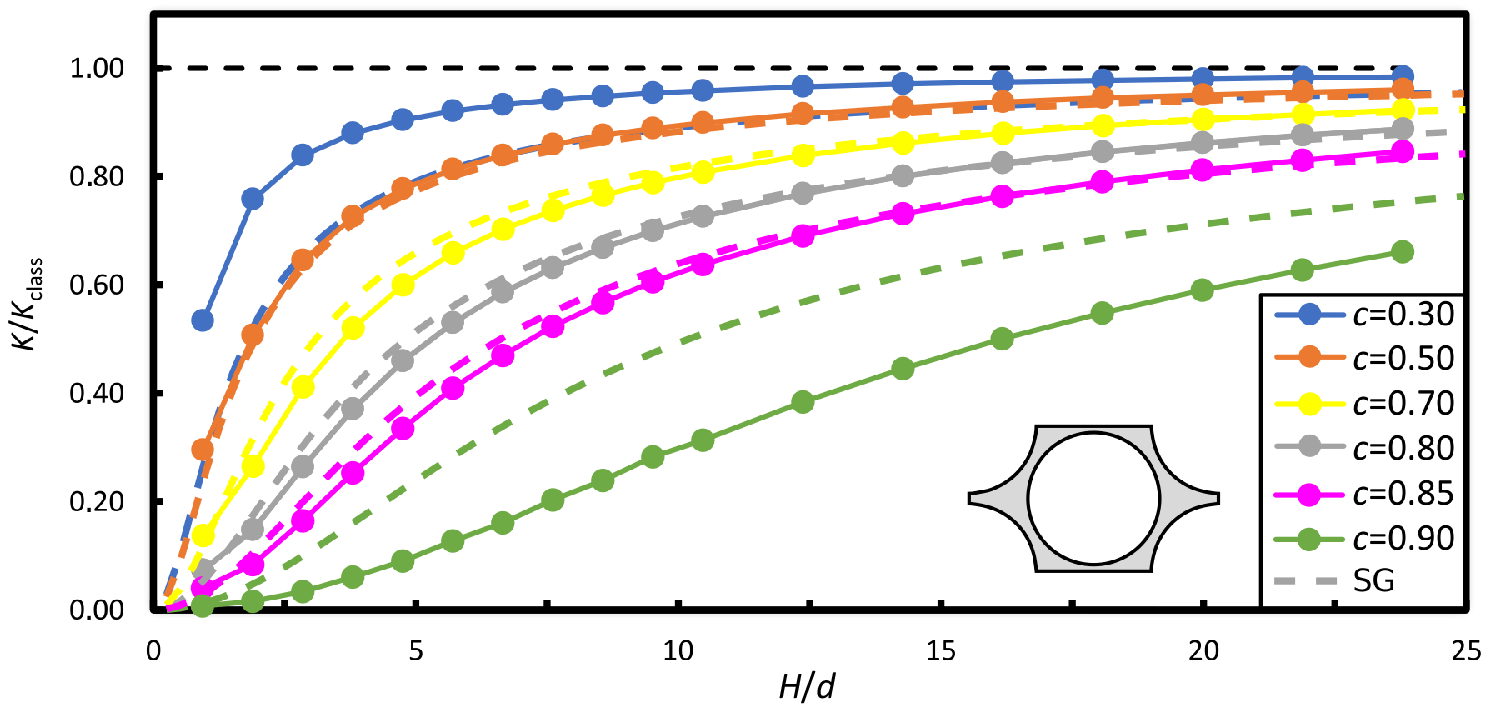}\label{fig:sizeeffectbending_circpore_orient2}}%
\caption[]{Size effects in bending in orientation 2 for \subref{fig:sizeeffectbending_hexpore_orient2} hexagonal and \subref{fig:sizeeffectbending_circpore_orient2} circular pores in comparison to predictions of stress-gradient theory (SG)}%
\label{fig:sizeeffectbending_orient2}%
\end{figure}

Figure~\ref{fig:sizeeffectbending_orient2} compares the size effects for hexagonal and circular pores in more detail and including higher porosities for orientation 2.
Again, hexagonal pores lead to a saturation of the size effect for $\VVF\gtrsim0.80$ (\figurename~\ref{fig:sizeeffectbending_hexpore_orient2}), whereas no such saturation is observed for circular pores in \figurename~\ref{fig:sizeeffectbending_circpore_orient2}.

Figure~\ref{fig:sizeeffectbending_circpore_orient2} incorporates additionally the respective prediction of the stress-gradient theory (Appendix~\ref{sec:bending_stress_gradient}) with higher-order constitutive parameters derived by homogenization with a composite-shell approach in \citep{Huetter2020}. It can be observed that the stress gradient theory provides very good predictions for medium porosities $0.5\leq\VVF\leq0.85$. {For a high porosity $\VVF=0.9$ the stress-gradient theory underestimates the size effect, presumably due to the employed composite shell approach which cannot reflect the pore interactions adequately for large porosities. For a lower porosity $\VVF=0.3$, the stress-gradient theory overestimates the size effect. Note that the composite-shell approach yields isotropic behavior and can thus not reflect the  anisotropy. Though, the stress-gradient theory itself requires a sixth-order tensor for the higher-order relation which could reflect an anisotropy in principle \citep{Auffray2009,Rizzi2019,Christensen1987}.}

\section{Summary and conclusions}
\label{sec:summary}

The present contribution employed a plane, regular hexagonal arrangement of pores as minimum models to investigate size effects of porous and cellular materials under simple shear, uniaxial loading and pure bending. Considering theoretically infinitely long specimens excludes potential coupling with effects of lateral surfaces. From a numerical point of view, this approach allows to resolve many more unit cells over the cross section than in previous studies from literature with specimens of finite length.   
{Firstly, it can be said that the present results comply qualitatively with the known trends from literature for (regular and disturbed) honeycomb structures, that positive size effects occur under simply shear \citep{Chen2002,Tekoglu2011,Liebenstein2018,Janicke2013,Iltchev2015,Rizzi2019a} and negative size under uniaxial loading or bending \citep{Tekoglu2011,Liebenstein2018}. This facts contradicts the prediction of \citet{Abdoul-Anziz2018} that honeycomb lattices with low relative density should behave purely classically. In the present contribution,}
models with circularly and hexagonally-shaped pores have been investigated and compared. The latter have prismatic struts, which have been modeled either as beams or as plane stress continua. 
Furthermore, two orientations of the lattice relative to the axes of loading have been investigated.
It was found that the classical effective elastic moduli are independent of the orientation as expected and that beam models are suitable for porosities larger than about 70\%.
But the size effects depend considerably on the alignment between lattice orientation and direction of loading. This means that the size effects are anisotropic, as can be anticipated from symmetry considerations in generalized continuum theories \citep{Auffray2009}.
In addition, the negative size effects under tension and bending, which originate from free surfaces, depend strongly on porosity and shape of the pores and thus on the material distribution along the struts. These size effects have been interpreted successfully by the stress-gradient theory, for which the bending solution was derived. 
Under simple shear, a positive (stiffening) size effect was observed which shows a significant anisotropy for hexagonal pores and thus prismatic struts.
The magnitudes of size effects in simple shearing are smaller in magnitude than for bending and uniaxial loading. In particular, it turned out that the rule of thumb that classical homogenization would be applicable when the ratio of scale separation $\height/\diamcell$ is larger than 10 \citep{Ameen2017} is not strict enough for bending and uniaxial loading at high porosities. A comparison of the present results from regular beam lattices with respective date from stochastic lattices from literature \citep{Tekoglu2011} showed that the size effects may be even stronger for the stochastic lattices.

With its systematic sensitivity study, the present contribution provides a comprehensive data base for the benchmark of generalized continuum theories. 
{Size effects in honeycomb structures have been modeled macroscopically by (first) strain-gradient \citep{Chen2002,Chung2009,Rosi2019,Rizzi2019}, Cosserat \citep{Tekoglu2011,Liebenstein2018,Huetter2019,Chung2009,Diebels2003,Glaesener2019}, (first order) micromorphic \citep{Janicke2013} or (first) stress-gradient theories \citep{Huetter2020}. First strain gradient and micromorphic theories, including the special case of Cosserat and couple stress theory, predict only positive size effects whereas the first stress gradient theory predicts only softening size effects. Thus, a second gradient theory or a second order micromorphic theory is necessary to describe the appearance of both type of effects for a single material under different loading conditions \citep{Forest2020}.
}
Future studies need to investigate the size effects for more realistic 3D models, potentially with a view on the rapidly evolving field of architectured materials.  
  
\bibliographystyle{elsarticle-harv}
\bibliography{FoamTopologySizeEffects}

\begin{thebibliography}{51}
\expandafter\ifx\csname natexlab\endcsname\relax\def\natexlab#1{#1}\fi
\expandafter\ifx\csname url\endcsname\relax
  \def\url#1{\texttt{#1}}\fi
\expandafter\ifx\csname urlprefix\endcsname\relax\def\urlprefix{URL }\fi

\bibitem[{Abdoul-Anziz and Seppecher(2018)}]{Abdoul-Anziz2018}
Abdoul-Anziz, H., Seppecher, P., 2018. Strain gradient and generalized continua
  obtained by homogenizing frame lattices. Mathematics and mechanics of complex
  systems 6~(3), 213--250.

\bibitem[{Ameen et~al.(2018{\natexlab{a}})Ameen, Peerlings, and
  Geers}]{Ameen2017}
Ameen, M.~M., Peerlings, R. H.~J., Geers, M. G.~D., 2018{\natexlab{a}}. A
  quantitative assessment of the scale separation limits of classical and
  higher-order asymptotic homogenization. Eur. J. Mech. A-Solid. 71, 89--100.

\bibitem[{Ameen et~al.(2018{\natexlab{b}})Ameen, Rokoš, Peerlings, and
  Geers}]{Ameen2018}
Ameen, M.~M., Rokoš, O., Peerlings, R. H.~J., Geers, M. G.~D.,
  2018{\natexlab{b}}. Size effects in nonlinear periodic materials exhibiting
  reversible pattern transformations. Mech. Mater. 124, 55--70.

\bibitem[{Anderson and Lakes(1994)}]{Anderson1994a}
Anderson, W.~B., Lakes, R.~S., 1994. Size effects due to cosserat elasticity
  and surface damage in closed-cell polymethacrylimide foam. J. Mater. Sci.
  29~(24), 6413--6419.

\bibitem[{Andrews et~al.(2001)Andrews, Gioux, Onck, and Gibson}]{Andrews2001}
Andrews, E.~W., Gioux, G., Onck, P., Gibson, L.~J., 2001. Size effects in
  ductile cellular solids. part ii: experimental results. Int. J. Mech. Sci.
  43~(3), 701--713.

\bibitem[{Ashby and Gibson(1997)}]{GibsonAshby}
Ashby, M.~F., Gibson, L.~J., 1997. Cellular solids: structure and properties.
  Press Syndicate of the University of Cambridge, Cambridge, UK, 175--231.

\bibitem[{Auffray et~al.(2009)Auffray, Bouchet, and Bréchet}]{Auffray2009}
Auffray, N., Bouchet, R., Bréchet, Y., 2009. Derivation of anisotropic matrix
  for bi-dimensional strain-gradient elasticity behavior. Int. J. Solids
  Struct. 46~(2), 440--454.

\bibitem[{Chen and Fleck(2002)}]{Chen2002}
Chen, C., Fleck, N.~A., 2002. Size effects in the constrained deformation of
  metallic foams. J. Mech. Phys. Solids. 50~(5), 955--977.

\bibitem[{Christensen(1987)}]{Christensen1987}
Christensen, R.~M., 1987. Sufficient symmetry conditions for isotropy of the
  elastic moduli tensor. J. Appl. Mech 54~(4), 772--777.

\bibitem[{Chung and Waas(2009)}]{Chung2009}
Chung, J., Waas, A.~M., 2009. The micropolar elasticity constants of circular
  cell honeycombs. Proc. R. Soc. A 465~(2101), 25--39.

\bibitem[{Diebels and Geringer(2014)}]{Diebels2014}
Diebels, S., Geringer, A., 2014. Micromechanical and macromechanical modelling
  of foams: Identification of {Cosserat} parameters. Z. Angew. Math. Mech.
  94~(5), 414--420.

\bibitem[{Diebels and Steeb(2003)}]{Diebels2003}
Diebels, S., Steeb, H., 2003. Stress and couple stress in foams. Comp. Mater.
  Sci. 28~(3-4), 714--722.

\bibitem[{Dillard et~al.(2006)Dillard, Forest, and Ienny}]{Dillard2006}
Dillard, T., Forest, S., Ienny, P., 2006. Micromorphic continuum modelling of
  the deformation and fracture behaviour of nickel foams. Eur. J. Mech.
  A-Solid. 25~(3), 526--549.

\bibitem[{Forest(2020)}]{Forest2020}
Forest, S., 2020. Strain gradient elasticity from capillarity to the mechanics
  of nano-objects. In: Mechanics of Strain Gradient Materials. Springer
  International Publishing, Cham, pp. 37--70.

\bibitem[{Forest and Sab(2012)}]{Forest2012}
Forest, S., Sab, K., 2012. Stress gradient continuum theory. Mech. Res. Commun.
  40, 16--25.

\bibitem[{Gauthier and Jahsman(1975)}]{Gauthier1975}
Gauthier, R.~D., Jahsman, W.~E., 1975. A quest for micropolar elastic
  constants. J. Appl. Mech. 42~(2), 369--374.

\bibitem[{Glaesener et~al.(2019)Glaesener, Lestringant, Telgen, and
  Kochmann}]{Glaesener2019}
Glaesener, R.~N., Lestringant, C., Telgen, B., Kochmann, D.~M., 2019. Continuum
  models for stretching- and bending-dominated periodic trusses undergoing
  finite deformations. Int. J. Solids Struct. 171, 117--134.

\bibitem[{Gong et~al.(2005)Gong, Kyriakides, and Jang}]{Gong2005}
Gong, L., Kyriakides, S., Jang, W.-Y., 2005. Compressive response of open-cell
  foams. part i: Morphology and elastic properties. Int. J. Solids Struct.
  42~(5), 1355--1379.

\bibitem[{Gross and Seelig(2006)}]{Gross2006}
Gross, D., Seelig, T., 2006. Fracture Mechanics -- With an Introduction to
  Micromechanics. Springer, Berlin Heidelberg.

\bibitem[{Ha et~al.(2016)Ha, Plesha, and Lakes}]{Ha2016}
Ha, C.~S., Plesha, M.~E., Lakes, R.~S., Jul. 2016. Chiral three-dimensional
  isotropic lattices with negative {Poisson}'s ratio. Phys. Status Solidi B
  253~(7), 1243--1251.

\bibitem[{Hashin(1962)}]{Hashin1962}
Hashin, Z., 1962. The elastic moduli of heterogeneous materials. J. Appl. Mech.
  29~(1), 143--150.

\bibitem[{Hård~af Segerstad et~al.(2009)Hård~af Segerstad, Toll, and
  Larsson}]{Haard2009}
Hård~af Segerstad, P., Toll, S., Larsson, R., Mar. 2009. A micropolar theory
  for the finite elasticity of open-cell cellular solids. Proc. R. Soc. A
  465~(2103), 843--865.

\bibitem[{Hütter(2016)}]{Huetter2016}
Hütter, G., 2016. Application of a microstrain continuum to size effects in
  bending and torsion of foams. Int. J. Eng. Sci. 101, 81--91.

\bibitem[{Hütter(2019{\natexlab{a}})}]{Huetter2019}
Hütter, G., 2019{\natexlab{a}}. On the micro-macro relation for the
  microdeformation in the homogenization towards micromorphic and micropolar
  continua. J. Mech. Phys. Solids 127, 62--79.

\bibitem[{Hütter(2019{\natexlab{b}})}]{HuetterHabil}
Hütter, G., 2019{\natexlab{b}}. A theory for the homogenisation towards
  micromorphic media and its application to size effects and damage.
  Habilitation, TU Bergakademie Freiberg.

\bibitem[{Hütter et~al.(2020)Hütter, Sab, and Forest}]{Huetter2020}
Hütter, G., Sab, K., Forest, S., 2020. Kinematics and constitutive relations
  in the stress-gradient theory: interpretation by homogenization. Int. J.
  Solids Struct. 193-194, 90--97.

\bibitem[{Iltchev et~al.(2015)Iltchev, Marcadon, Kruch, and
  Forest}]{Iltchev2015}
Iltchev, A., Marcadon, V., Kruch, S., Forest, S., 2015. Computational
  homogenisation of periodic cellular materials: Application to structural
  modelling. Int. J. Mech. Sci. 93, 240--255.

\bibitem[{Jang et~al.(2008)Jang, Kraynik, and Kyriakides}]{Jang2008}
Jang, W.-Y., Kraynik, A.~M., Kyriakides, S., 2008. On the microstructure of
  open-cell foams and its effect on elastic properties. Int. J. Solids Struct.
  45~(7), 1845--1875.

\bibitem[{Jänicke et~al.(2013)Jänicke, Sehlhorst, Duster, and
  Diebels}]{Janicke2013}
Jänicke, R., Sehlhorst, H.-G., Duster, A., Diebels, S., 2013. Micromorphic
  two-scale modelling of periodic grid structures. Int. J. Multiscale Com.
  11~(2), 161--176.

\bibitem[{Kaiser et~al.(2020)Kaiser, Forest, and Menzel}]{Kaiser2020}
Kaiser, T., Forest, S., Menzel, A., 2020. A finite element implementation of
  the stress gradient theory. Meccanica, DOI:10.1007/s11012-020-01266-3.

\bibitem[{Lee et~al.(1996)Lee, Choi, and Choi}]{Lee1996}
Lee, J., Choi, J., Choi, K., 1996. Application of homogenization fem analysis
  to regular and re-entrant honeycomb structures. J. Mater. Sci. 31~(15),
  4105--4110.

\bibitem[{Liebenstein et~al.(2018)Liebenstein, Sandfeld, and
  Zaiser}]{Liebenstein2018}
Liebenstein, S., Sandfeld, S., Zaiser, M., 2018. Size and disorder effects in
  elasticity of cellular structures: From discrete models to continuum
  representations. Int. J. Solids Struct. 146, 97--116.

\bibitem[{Liebold and Müller(2016)}]{Liebold2016a}
Liebold, C., Müller, W.~H., 2016. Applications of higher-order continua to
  size effects in bending: Theory and recent experimental results. In:
  Generalized Continua as Models for Classical and Advanced Materials. Springer
  International Publishing, Cham, pp. 237--260.

\bibitem[{Liu and Su(2009)}]{Liu2009}
Liu, S., Su, W., 2009. Effective couple-stress continuum model of cellular
  solids and size effects analysis. Int. J. Solids. Struct. 46~(14--15),
  2787--2799.

\bibitem[{Mühlich(2020)}]{Muehlich2020}
Mühlich, U., 2020. Deformation and failure onset of random elastic beam
  networks generated from the same type of random graph. In: Developments and
  Novel Approaches in Biomechanics and Metamaterials. Springer International
  Publishing, Cham, pp. 393--408.

\bibitem[{Redenbach et~al.(2012)Redenbach, Shklyar, and Andrä}]{Redenbach2012}
Redenbach, C., Shklyar, I., Andrä, H., 2012. Laguerre tessellations for
  elastic stiffness simulations of closed foams with strongly varying cell
  sizes. Int. J. Eng. Sci. 50~(1), 70--78.

\bibitem[{Rizzi et~al.(2019{\natexlab{a}})Rizzi, Dal~Corso, Veber, and
  Bigoni}]{Rizzi2019}
Rizzi, G., Dal~Corso, F., Veber, D., Bigoni, D., 2019{\natexlab{a}}.
  Identification of second-gradient elastic materials from planar hexagonal
  lattices. part i: Analytical derivation of equivalent constitutive tensors.
  Int. J. Solids Struct. 176-177, 1--18.

\bibitem[{Rizzi et~al.(2019{\natexlab{b}})Rizzi, Dal~Corso, Veber, and
  Bigoni}]{Rizzi2019a}
Rizzi, G., Dal~Corso, F., Veber, D., Bigoni, D., 2019{\natexlab{b}}.
  Identification of second-gradient elastic materials from planar hexagonal
  lattices. part ii: Mechanical characteristics and model validation. Int. J.
  Solids Struct. 176-177, 19--35.

\bibitem[{Rosi and Auffray(2019)}]{Rosi2019}
Rosi, G., Auffray, N., 2019. Continuum modelling of frequency dependent
  acoustic beam focussing and steering in hexagonal lattices. Eur. J. Mech.
  A-Solid. 77, 103803.

\bibitem[{Rueger and Lakes(2016)}]{Rueger2016}
Rueger, Z., Lakes, R.~S., Jan. 2016. Experimental {Cosserat} elasticity in
  open-cell polymer foam. Philos. Mag. 96~(2), 93--111.

\bibitem[{Soyarslan et~al.(2019)Soyarslan, Blümer, and
  Bargmann}]{Soyarslan2019}
Soyarslan, C., Blümer, V., Bargmann, S., 2019. Tunable auxeticity and
  elastomechanical symmetry in a class of very low density core-shell cubic
  crystals. Acta. Mater. 177, 280--292.

\bibitem[{Storm et~al.(2019)Storm, Abendroth, and Kuna}]{Storm2019}
Storm, J., Abendroth, M., Kuna, M., 2019. Effect of morphology, topology and
  anisoptropy of open cell foams on their yield surface. Mech. Mater. 137,
  103145.

\bibitem[{Storm et~al.(2013)Storm, Abendroth, Zhang, and Kuna}]{Storm2013}
Storm, J., Abendroth, M., Zhang, D., Kuna, M., 2013. Geometry dependent
  effective elastic properties of open cell foams based on {Kelvin} cell
  models. Adv. Eng. Mater. 15~(12), 1292--1298.

\bibitem[{Tekoğlu et~al.(2011)Tekoğlu, Gibson, Pardoen, and
  Onck}]{Tekoglu2011}
Tekoğlu, C., Gibson, L., Pardoen, T., Onck, P., 2011. Size effects in foams:
  Experiments and modeling. Prog. Mater. Sci. 56~(2), 109--138.

\bibitem[{Tekoğlu and Onck(2008)}]{Tekoglu2008}
Tekoğlu, C., Onck, P.~R., 2008. Size effects in two-dimensional {Voronoi}
  foams: A comparison between generalized continua and discrete models. J.
  Mech. Phys. Solids. 56~(12), 3541--3564.

\bibitem[{Warren and Kraynik(1987)}]{Warren1987}
Warren, W., Kraynik, A., 1987. Foam mechanics: the linear elastic response of
  two-dimensional spatially periodic cellular materials. Mech. Mater. 6~(1),
  27--37.

\bibitem[{Waseem et~al.(2013)Waseem, Beveridge, Wheel, and Nash}]{Waseem2013}
Waseem, A., Beveridge, A., Wheel, M., Nash, D., 2013. The influence of void
  size on the micropolar constitutive properties of model heterogeneous
  materials. Eur. J. Mech. A-Solid. 40, 148--157.

\bibitem[{Wheel et~al.(2015)Wheel, Frame, and Riches}]{Wheel2015}
Wheel, M.~A., Frame, J.~C., Riches, P.~E., 2015. Is smaller always stiffer? on
  size effects in supposedly generalised continua. Int. J. Solids. Struct.
  67-68, 84--92.

\bibitem[{Yang and Lakes(1982)}]{Yang1982}
Yang, J., Lakes, R.~S., 1982. Experimental study of micropolar and couple
  stress elasticity in compact bone in bending. J. Biomech. 15~(2), 91--98.

\bibitem[{Yoder et~al.(2019{\natexlab{a}})Yoder, Thompson, and
  Summers}]{Yoder2019}
Yoder, M., Thompson, L., Summers, J., 2019{\natexlab{a}}. Size effects in
  lattice-structured cellular materials: edge softening effects. J. Mater. Sci.
  54~(5), 3942--3959.

\bibitem[{Yoder et~al.(2019{\natexlab{b}})Yoder, Thompson, and
  Summers}]{Yoder2019a}
Yoder, M., Thompson, L., Summers, J., 2019{\natexlab{b}}. Size effects in
  lattice-structured cellular materials: material distribution. J. Mater. Sci.
  54~(18), 11858--11877.

\end{thebibliography}

\appendix

\section{Composite Shell Homogenization for the Plane Case}
\label{sec:composite shell}

The effective moduli (in-plane bulk modulus and shear modulus) of materials with circular pores (or inclusions) have been determined in \citep{Huetter2019} for kinematic boundary conditions and plane stresses as
\begin{align}
   \bulkmodeff&
	             =\emod\frac{1}{2}\frac{1-\VVF}{1-\Poissrat+\VVF(1+\Poissrat)} \label{eq:effbulkmodcirc}\\
	\shearmodeff&
	=\shearmod \frac{\left(4-(1+\Poissrat)(1-\VVF)^3 \right)\left(1-\VVF\right)}{(3-\Poissrat)(1+\Poissrat+(3-4\Poissrat)\VVF+(1+\Poissrat)\VVF^4)+(1+\Poissrat)^2\VVF(3-6\VVF+4\VVF^2)}
\label{eq:effmodcirc}
\end{align}
The effective Young's modulus follows as $\emodeff=4\shearmodeff\bulkmodeff/(\shearmodeff+\bulkmodeff)$. The corresponding effective shear modulus under statically uniform boundary conditions was determined in \citep{HuetterHabil} as
\begin{align}
	\shearmodeff=\shearmod\,\frac{\left(1-\VVF\right)^3(1+\Poissrat)}{1+\Poissrat+(1-3\Poissrat)\VVF+(7+3\Poissrat)\VVF^2+(3-\Poissrat)\VVF^3}\,.
	\label{eq:effshearmodstatBC}
\end{align}
The in-plane bulk modulus $\bulkmodeff$ in eq.~\eqref{eq:effbulkmodcirc} is the same for both types of boundary conditions.

\section{Bending Solution in Stress-Gradient Theory}
\label{sec:bending_stress_gradient}

The stress-gradient theory of \citet{Forest2012} introduces the field of \enquote{micro-displacements} $\microdisplcomp_{ijk}$  as work-conjugate quantity to the gradient 
$\stressgradcomp_{ijk}:=\partial\stresscomp_{ij}/\partial x_k$ of the Cauchy stress $\stresscomp_{ij}$. The latter satisfies the conventional balances $\partial\stresscomp_{ij}/\partial x_i=0$ and $\stresscomp_{ij}=\stresscomp_{ji}$. Energetic considerations imply that the kinematic relation 
\begin{equation}
	\straincomp_{ij}=\frac12\left(\frac{\partial\displcomp_{i}}{\partial x_j}+\frac{\partial\displcomp_{j}}{\partial x_i}\right)+\frac{\partial\microdisplcomp_{ijk}}{\partial x_k}
	\label{eq:strainSG}
\end{equation}
for the strain tensor $\straincomp_{ij}$ must not involve only the symmetric part of the gradient of the conventional displacement vector $\displcomp_{i}$, but additionally the divergence of micro-displacement tensor $\microdisplcomp_{ijk}$. The linear-elastic non-classical constitutive relations for an isotropic material\footnote{The linear-elastic constitutive relation between $\stressgradcomp_{ijk}$ and $\microdisplcomp_{kmp}$ is formed in general by a tensor of sixth-order tensor, which may be anisotropic for materials with hexagonal structure \citep{Auffray2009}.} read for the plane case in a Voigt-type notation
\begin{align}
\begin{pmatrix}
	3\microdisplcomp_{111}\\
	\microdisplcomp_{221}
\end{pmatrix}
&=\complmatSG\cdot
\begin{pmatrix}
	\stressgradcomp_{111}\\
	\stressgradcomp_{221}
\end{pmatrix}
\,,&
\begin{pmatrix}
	3\microdisplcomp_{222}\\
	\microdisplcomp_{112}
\end{pmatrix}
&=\complmatSG\cdot
\begin{pmatrix}
	\stressgradcomp_{222}\\
	\stressgradcomp_{112}
\end{pmatrix}\,.
\label{eq:linearelastcompliance}
\end{align}
Further details like symmetries of $\stressgradcomp_{ijk}$ and $\microdisplcomp_{ijk}$ can be found in \citep{Forest2012}.
The symmetric higher-order compliance matrix $\complmatSG$ was identified for a porous material by homogenization in \citep{Huetter2020}.

\begin{figure}[bht]
	\centering
	\inputsvg{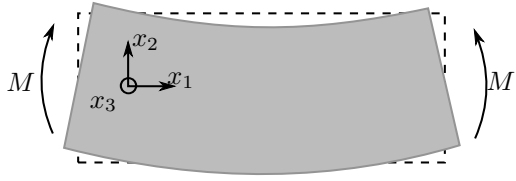}
	\caption{Pure bending}
	\label{fig:bendinghomogen}
\end{figure}

\begin{figure}
	\centering
	\subfloat[]{\inputgnuplot{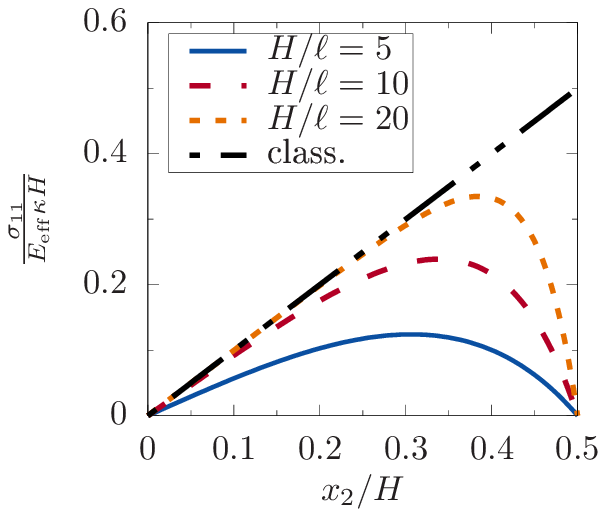}\label{fig:SGbendstress}}
	\hspace{1cm}
	\subfloat[]{\inputgnuplot{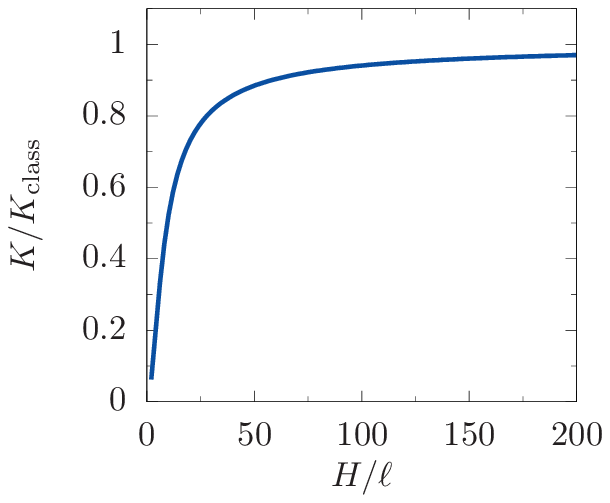}\label{fig:size_effect_SG_bend}}
	\caption[]{Size effect under bending predicted by stress-gradient theory in comparison to classical solution: \subref{fig:SGbendstress} stress field and \subref{fig:size_effect_SG_bend} bending stiffness}
\end{figure}

The stress-gradient solution for the pure bending problem in \figurename~\ref{fig:bendinghomogen}
is obtained by the ansatz that the only non-vanishing stress component $\stresscomp_{11}=\stresscomp_{11}(x_2)$ depends only on the distance $x_2$ to the neutral axis. Hooke's law for plane stresses thus yields $\straincomp_{11}=\stresscomp_{11}(x_2)/\emodeff$ and $\straincomp_{22}=-\Poissrat\stresscomp_{11}(x_2)/\emodeff$. Inserting the only non-vanishing component of the stress gradient $\stressgradcomp_{112}=\stresscomp'_{11}(x_2)$ to the higher-order constitutive law~\eqref{eq:linearelastcompliance}$_2$ yields $\microdisplcomp_{112}=\complmatSGcomp_{22}\stresscomp'_{11}(x_2)$ and $\microdisplcomp_{222}=1/3\complmatSGcomp_{12}\stresscomp'_{11}(x_2)$. These micro-displacements and strains can be inserted into the kinematic relation~\eqref{eq:strainSG} to obtain the ODEs
\begin{align}
	\stresscomp_{11}-\emodeff\complmatSGcomp_{22}\stresscomp_{11}''&=\frac{\partial\displcomp_{1}}{\partial x_1},&
	\Poissrat\stresscomp_{11}+\frac13\emodeff\complmatSGcomp_{12}\stresscomp_{11}''&=\frac{\partial\displcomp_{2}}{\partial x_2}\,.
	\label{eq:ODEbendSG}
\end{align}
It is known from micromorphic theories \citep{Gauthier1975,Huetter2016} that the Bernoulli hypothesis remains valid and non-classical terms affect only the lateral contraction $u_{\mathrm{lat}}$. Thus, we choose $\partial\displcomp_{1}/\partial x_1=\curvature x_2$ and $\partial\displcomp_{2}/\partial x_2= u_{\mathrm{lat}}'(x_2)$ with curvature $\curvature$=const (compare \eqref{eq:BC_bending}). The first ODE from \eqref{eq:ODEbendSG} thus yields
\begin{equation}
	\stresscomp_{11}=\emodeff\curvature\left[x_2-\frac{\height\sinh(\frac{x_2}{\charlengthSGtens})}{2\sinh(\frac{\height}{2\charlengthSGtens})}\right]\,.
	\label{eq:stressbendSG}
\end{equation}
Therein, $\charlengthSGtens=\sqrt{\emodeff\complmatSGcomp_{22}}$ abbreviates an internal characteristic length scale (identical to uniaxial loading in \citep{Huetter2020}). The constants of integration in \eqref{eq:stressbendSG} have been determined by the trivial boundary condition $\stresscomp_{11}(\pm \height/2)=0$. {A plot of the stress distribution in \figurename~\ref{fig:SGbendstress} exhibits similarities with the torsion solution in \citep{Kaiser2020}.}
The second equation \eqref{eq:ODEbendSG}$_2$ would now allow to determine the lateral contraction $u_{\mathrm{lat}}(x_2)$. Finally, the bending stiffness $\bendstiffness$ is obtained from the total strain-energy in each cross section
\begin{equation}
	\int\limits_{-\height/2}^{\height/2}\frac{1}{2}\left[\stresscomp_{ij}\straincomp_{ij}+\stressgradcomp_{ijk}\microdisplcomp_{ijk}\right]\,\mathrm{d}x_2
=\frac{1}{2}\int\limits_{-\height/2}^{\height/2}\left[\frac{1}{\emodeff}\stresscomp_{11}^2+\complmatSGcomp_{22}\left(\stresscomp'_{11}\right)^2\right]\,\mathrm{d}x_2
=\frac{1}{2}\bendstiffness\curvature^2
\end{equation}
as
\begin{equation}
	\bendstiffness=\frac{\emodeff\height^3}{12}\left[1-12\frac{\charlengthSGtens^2}{\height^2}\left(\frac{\frac{\height}{2\charlengthSGtens}}{\tanh\left(\frac{\height}{2\charlengthSGtens}\right)}-1\right)\right]\,.
\end{equation}
Obviously, the term in the square bracket reflects the size effect. {Its plot in \figurename~\ref{fig:size_effect_SG_bend} shows that the stress-gradient theory predicts a softening size effect as known from uniaxial loading and torsion \citep{Huetter2020,Kaiser2020}.}

\end{document}